\documentclass[twocolumn, secnumarabic, amssymb, superscriptaddress, aps, prl,nobibnotes, longbibliography]{revtex4-2}
\usepackage{graphicx}
\usepackage{amsmath}
\usepackage{upgreek}
\usepackage{multirow}
\usepackage{booktabs}
\usepackage{array}
\usepackage{color}
\usepackage{xcolor}
\definecolor{navy}{rgb}{0,0,0.64}
\usepackage{setspace}
\usepackage{hyperref}
\hypersetup{colorlinks,breaklinks,
            urlcolor=[rgb]{0,0,0.64},
            linkcolor=[rgb]{0,0,0.64},
            citecolor=[rgb]{0,0,0.64}}
\usepackage{nccmath}
\usepackage{amssymb}
\usepackage{bm}
\usepackage{lineno}

\usepackage{tikz}
\newcommand*\circled[1]{\tikz[baseline=(char.base)]{
            \node[shape=circle,draw,inner sep=0.4pt] (char) {#1};}}

\setcitestyle{super}

\def\bibsection{\section{R\lowercase{eferences}}} 

\newcommand{\beginsupplement}{%
        \setcounter{secnumdepth}{1} 
        \setcounter{table}{0}
        \renewcommand{\thetable}{S\arabic{table}}%
        \setcounter{figure}{0}
        \renewcommand{\thefigure}{S\arabic{figure}}%
        \setcounter{equation}{0}
        \renewcommand{\theequation}{S\arabic{equation}}%
        \setcounter{section}{0}
        \renewcommand\thesection{Supplementary Note~\arabic{section}}%
     }

\newcommand{\beginSuppRef}{%
        \setcounter{secnumdepth}{0} 
        \setcounter{section}{0}
        \renewcommand\thesection{}%
     }

\newcommand{\papertitle}{Core-level signature of long-range density-wave order and short-range excitonic correlations probed by attosecond broadband spectroscopy}

\newcommand{\Cal}{University of California at Berkeley, Department of Chemistry, Berkeley, CA 94720, USA}
\newcommand{\LBNL}{Materials Sciences Division, Lawrence Berkeley National Laboratory, Berkeley, CA 94720, USA}
\newcommand{\SPST}{School of Physical Science and Technology, ShanghaiTech University, Shanghai 201210, China}
\newcommand{\STLTP}{ShanghaiTech Laboratory for Topological Physics, ShanghaiTech University, Shanghai 201210, China}
\newcommand{\TDLI}{Tsung-Dao Lee Institute and Zhangjiang Institute for Advanced Study, Shanghai Jiao Tong University, Shanghai 201210, China}
\newcommand{\KLLP}{Key Laboratory for Laser Plasmas (Ministry of Education), School of Physics and Astronomy, Shanghai Jiao Tong University, Shanghai 200240, China}
\newcommand{\CCS}{Center for Computational Sciences, University of Tsukuba, Tsukuba, Ibaraki 305-8577, Japan}
\newcommand{\Max}{Max Planck Institute for the Structure and Dynamics of Matter, Luruper Chaussee 149, 22761 Hamburg, Germany}

\begin{document}

\title{\papertitle}

\author{Alfred~Zong}
\thanks{These authors contributed equally to this work: Alfred~Zong and Sheng-Chih~Lin.}
\affiliation{\Cal}
\affiliation{\LBNL}

\author{Sheng-Chih~Lin}
\thanks{These authors contributed equally to this work: Alfred~Zong and Sheng-Chih~Lin.}
\affiliation{\Cal}
\affiliation{\LBNL}

\author{Shunsuke~A.~Sato}
\affiliation{\CCS}
\affiliation{\Max}

\author{Emma~Berger}
\affiliation{\Cal}

\author{Bailey~R.~Nebgen}
\affiliation{\Cal}
\affiliation{\LBNL}

\author{Marcus~Hui}
\affiliation{\Cal}

\author{B.~Q.~Lv}
\affiliation{\TDLI}

\author{Yun~Cheng}
\affiliation{\KLLP}
\affiliation{\TDLI}

\author{Wei~Xia}
\affiliation{\SPST}
\affiliation{\STLTP}

\author{Yanfeng~Guo}
\affiliation{\SPST}
\affiliation{\STLTP}

\author{Dao~Xiang}
\affiliation{\KLLP}
\affiliation{\TDLI}

\author{Michael~W.~Zuerch}
\email[Correspondence to: ]{mwz@berkeley.edu}
\affiliation{\Cal}
\affiliation{\LBNL}

\begin{abstract}
    Advances in attosecond core-level spectroscopies have successfully unlocked the fastest dynamics involving high-energy electrons. Yet, these techniques are not conventionally regarded as an appropriate probe for low-energy quasiparticle interactions that govern the ground state of quantum materials, nor for studying long-range order because of their limited sensitivity to local charge environments. Here, by employing a unique cryogenic attosecond beamline, we identified clear core-level signatures of long-range charge-density-wave formation in a quasi-2D excitonic insulator candidate, even though equilibrium photoemission and absorption measurements of the same core levels showed no spectroscopic singularity at the phase transition. Leveraging the high time resolution and intrinsic sensitivity to short-range charge excitations in attosecond core-level absorption, we observed compelling time-domain evidence for excitonic correlations in the normal-state of the material, whose presence has been subjected to a long-standing debate in equilibrium experiments because of interfering phonon fluctuations in a similar part of the phase space. Our findings support the scenario that short-range excitonic fluctuations prelude long-range order formation in the ground state, providing important insights in the mechanism of exciton condensation in a quasi-low-dimensional system. These results further demonstrate the importance of a simultaneous access to long- and short-range order with underlying dynamical processes spanning a multitude of time- and energy-scales, making attosecond spectroscopy an indispensable tool for both understanding the equilibrium phase diagram and for discovering novel, nonequilibrium states in strongly correlated materials.
\end{abstract}

\date{\today}

\maketitle

\section{I\lowercase{ntroduction}\label{sec:intro}}

The discovery of emerging nonequilibrium states in quantum materials is greatly advanced by the development of time-resolved spectroscopies that can capture the energetic fingerprints of various excitations and microscopic interactions. For strongly-correlated materials where elementary processes such as screening, charge transfer, and electron-electron scattering play an important role in determining their nonequilibrium properties, techniques with sub- to few-femtosecond temporal resolutions are necessary to examine these processes at their intrinsic timescales \cite{Petek1997,Bovensiepen2012,Zong2023}. Considering the fundamental limit imposed by the time-energy uncertainty principle where a short pulse necessitates a large bandwidth, to access such a short timescale, one has to resort to time-resolved core-level spectroscopies that utilize photons with energies in the extreme-ultraviolet (XUV) to soft X-ray regime. Although there has been tremendous innovation in building such photon sources and gaining attosecond-level information via either table-top setups \cite{Krausz2009,Lloyd-Hughes2021,Zong2023} or free-electron lasers \cite{Duris2020,Guo2024}, there remains considerable doubt on whether attosecond core-level techniques can be simultaneously sensitive to the low-energy processes, which are closely intertwined with the fast, high-energy dynamics in a correlated system \cite{Anderson2007,Morosan2012,Phillips2022,Alexandradinata2022}. Moreover, the sensitivity to local charge environments in core-level spectroscopies calls for questions on whether such probes can unambiguously distinguish long-range order from short-range correlations, the latter of which are often present in low-dimensional materials before a thermodynamic transition takes place.

To bridge the gap between high-energy, attosecond processes and low-energy quasiparticle interactions that drive phase transitions, we turn to attosecond broadband XUV absorption spectroscopy (ABXAS) \cite{Krausz2009,Zong2023}, which, in principle, covers all time, energy, and length scales discussed above. With a bandwidth that can nowadays span tens to hundreds of eV (ref.~\cite{Sidiropoulos2021}), it is expected to provide sub-femtosecond temporal resolution while being sensitive to meV-level changes in the absorption features caused by low-energy processes \cite{Geneaux2019}. We would also like to emphasize the broadband and continuous spectral coverage of ABXAS (Fig.~\ref{fig:HHG_source})
as it is critical to assess subtle low-energy changes in different spectral windows that are sensitive to different underlying dynamics. To the lowest order, absorption here involves excitation of localized core-level electrons to empty itinerant valence states, so the resulting spectra are expected to concurrently capture both local and non-local evolutions of the electronic structure. 

Here, we apply ABXAS to examine 1$T$-TiSe$_2$, a layered charge-density-wave (CDW) compound that forms a $2\times2\times2$ superlattice below a transition temperature $T_c \approx 200$~K, a process that is putatively accompanied by exciton condensation at the same temperature \cite{Wilson1977, Rossnagel2002,Cercellier2007,Kogar2017,Lian2020}. Due to the strong-coupling nature of the CDW \cite{Rossnagel2011,Watson2020}, considerable CDW fluctuations are found above $T_c$ in the form of short-range $2\times2$ order in each layer \cite{Chen2016,Cheng2022}, providing an ideal platform to benchmark the effect of local vs. extended orders in ABXAS. Despite decades of close scrutiny, whether and how excitonic correlations drive the CDW transition remains under debate, and one of the most convincing pieces of evidence so far that reports on the collective excitation of the exciton condensation \cite{Kogar2017} has recently come into question \cite{Lian2019,Lin2022}. In this work, we show that not only does ABXAS demonstrate clear core-level signatures about the formation of long-range CDW --- which are absent in static core-level spectroscopies --- its superior temporal resolution and sensitivity to local charge dynamics also help reveal persistent short-range excitonic fluctuations well above $T_c$, which are hard to be distinguished from phononic fluctuations in equilibrium. Our findings show that excitonic correlations play a key role in the buildup of local CDW amplitude before the thermodynamic transition occurs, offering the missing piece about the excitonic character of the ground state in this quasi-2D material where fluctuations are expected to abound.

\begin{figure*}[htb!]
	\includegraphics[width=0.99\textwidth]{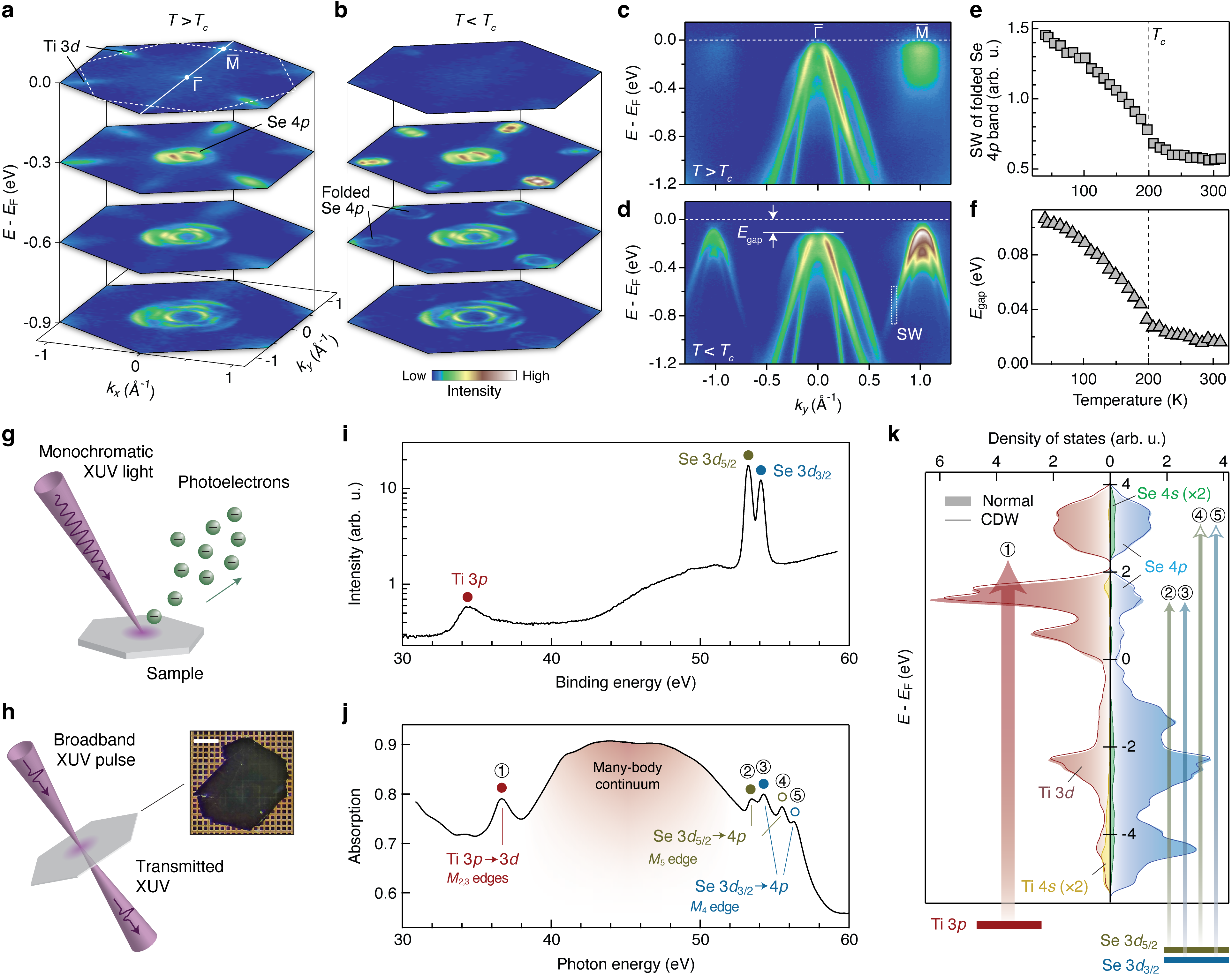}
	\caption{\textbf{Valence and core electronic structures of 1\textit{T}-TiSe$_\text{2}$ in equilibrium.} \textbf{a},\textbf{b},~Constant energy cuts of photoemission intensity at 260~K ($T>T_c$,~\textbf{a}) and 40~K ($T<T_c$,~\textbf{b}). In the normal state, the Fermi surface is dominated by electron-like Ti~3$d$ band at $\bar{M}$ and hole-like Se~4$p$ bands at $\bar{\Gamma}$. Due to the superlattice formation below $T_c$, folded Se~4$p$ bands are clearly visible at the $\bar{M}$ points in \textbf{b}. \textbf{c},\textbf{d},~Dispersions along $\bar{M}$-$\bar{\Gamma}$-$\bar{M}$ (white line in \textbf{a}) above and below $T_c$, highlighting the gap formation $E_\text{gap}$ in addition to the replica Se~4$p$ bands in the CDW ground state. \textbf{e},~Temperature-dependent spectral weight (SW) of the folded Se~4$p$ bands taken from the dashed rectangle in \textbf{d}. \textbf{f},~Temperature-dependent gap size $E_\text{gap}$ at $\bar{\Gamma}$, as labeled in \textbf{d}. $E_\text{gap}$ is calculated as the energy difference between the chemical potential and the leading edge of the energy distribution curves (see Fig.~\ref{fig:static_valence}b). \textbf{g},\textbf{h},~Schematic of the two techniques used to probe the core-level electronic states: synchrotron-based photoemission (\textbf{g}) and absorption spectroscopy based on a table-top broadband attosecond XUV source (\textbf{h}). An optical image of the freestanding, single-crystalline flake used in the absorption measurement is shown in \textbf{h}; scale bar:~200~$\upmu$m. \textbf{i},\textbf{j},~Core-level photoemission and absorption spectra, respectively, in the 30--60~eV window measured above $T_c$. Labeled peaks {\protect\circled{1}}--{\protect\circled{5}} in \textbf{j} correspond to the vertical arrows in \textbf{k}. See Fig.~\ref{fig:static_core} for temperature-dependent static core-level photoemission and absorption spectra, where no signature about the phase transition was detected. \textbf{k},~Orbital-projected density of states near the Fermi level for both normal (shaded) and CDW phase (solid curve). Data for Ti~4$s$ and Se~4$s$ orbitals are multiplied by 2 for clearer visualization. Experimentally observed core-to-valence transitions are labeled.}
\label{fig:1}
\end{figure*}

\section{A\lowercase{bsence of phase transition signal in static core-level spectroscopies}}

We start by characterizing the static CDW transition using both core-level photoemission and XUV absorption to look for spectroscopic signatures, if any, during the equilibrium phase transition. In the photoemission experiment (see \ref{sn:arpes}), we simultaneously measured the momentum-resolved spectra near the Fermi level, where the band structure is known to undergo significant renormalization \cite{Rossnagel2011} when the long-range order sets in at $T_c$. Figures~\ref{fig:1}a and \ref{fig:1}b show a series of constant-energy cuts of the photoemission intensity above and below $T_c$, respectively. Due to the $2\times2\times2$ reconstruction, the hole-like bands near the $\bar{\Gamma}$ point with mostly Se~4$p$ character are folded into six $\bar{M}$ points, as labeled in Fig.~\ref{fig:1}b. The appearance of the CDW-induced band replica is further evident in the dispersion map along the $\bar{M}$-$\bar{\Gamma}$-$\bar{M}$ direction (Fig.~\ref{fig:1}d), which is accompanied by an energy gap $E_\text{gap}$ that is most clearly resolved at $\bar{\Gamma}$. Both the energy gap and the replica band spectral weight display an order-parameter-like onset at $T_c$ (Fig.~\ref{fig:1}e,f). Above $T_c$, neither quantity is a constant function of temperature but instead decreases slightly with an increasing temperature, suggesting that short-range fluctuations are present and the phase transition deviates from a mean-field description. The fluctuating character of the order parameter is particularly pronounced as we examine the replica band spectral weight above $T_c$ (Fig.~\ref{fig:1}a,c): at the $\bar{M}$ point below the Fermi level, incoherent photoemission intensity resulting from the folded Se~4$p$ bands is clearly observed above the background, which is a consequence of the short-range 2D CDW order, or equivalently, a soft transverse optical phonon responsible for the CDW formation \cite{Holt2001}. However, given the intertwined nature of the lattice distortion and excitonic correlations in this material \cite{Wilson1977, Rossnagel2002,Cercellier2007,VanWezel2010,Kogar2017,Lian2020,Cheng2022}, the question remains on whether the short-range order is purely phononic in nature or whether it also contains contributions from the excitonic interaction. Utilizing the sensitivity to local charge density in core-level spectroscopies, we will return to addressing this question using ABXAS.

Unlike the valence electronic states that display a clear change at $T_c$, such a signature of long-range order formation is absent in core-level spectroscopies, including both core-level photoemission and absorption measurements (Fig.~\ref{fig:1}g,h). For the core levels, shown in Fig.~\ref{fig:1}i, we focused on the broad Ti~3$p$ peak at 34.35~eV binding energy and the relatively sharp Se~3$d$ peaks around 53--54~eV, the latter of which form a doublet due to spin-orbit coupling. These peaks were selected because they yield strong dipole-allowed core-to-valence XUV absorption features where the valence Se~4$p$ and Ti~3$d$ states directly participate in the CDW formation. A schematic of the relevant transitions is shown in Fig.~\ref{fig:1}k for both Ti and Se states that are projected to their respective orbitals (see \ref{sn:dft} for computational details). For Se, four absorption peaks are observed, labeled \circled{2} to \circled{5} in Fig.~\ref{fig:1}j, which correspond to the $M_{4,5}$ transitions between the two spin-orbit-split 3$d$ core levels and the two clusters of Se~4$p$ states above the Fermi level $E_F$ (Fig.~\ref{fig:1}k): one cluster spans from 0 to 2~eV, the second one from 2.3 to 4~eV, all measured relative to $E_F$. Considering the binding energies of the Se~3$d$ core levels observed in photoemission (Fig.~\ref{fig:1}i), the slight blue shift of the Se peak positions in the absorption spectra almost exactly reflects the energy of the two clusters of unoccupied Se~4$p$ states right above $E_F$. On the other hand, the position of the Ti~$M_{2,3}$ absorption edges corresponding to the $3p\rightarrow3d$ transition (labeled as \circled{1} in Fig.~\ref{fig:1}j,k) is blue-shifted by more than 2~eV compared to the binding energy of the Ti~3$p$ core position, suggesting a breakdown of the single-particle description of the absorption process. Indeed, from our time-dependent density functional theory calculations (see Fig.~\ref{fig:abs_cal} and \ref{sn:TDDFT}), the computed absorption spectra look drastically different between the complete calculation and the independent particle approximation where the Hartree and exchange-correlation terms are excluded in the time propagation. Notably, besides the spectral blue shift, a broad absorption peak ranging from 38 to 53~eV in the measured Ti spectrum (Fig.~\ref{fig:1}j) is correctly reproduced in the complete calculation but absent in the independent particle approximation (Fig.~\ref{fig:abs_cal}b). Following the assignment of a previous static absorption measurement \cite{Heinrich2023}, we interpret this broad feature as a many-body continuum peak due to valence-electron excitations in addition to the core-to-valence transition. We hence expect this broad spectral range of Ti to be particularly sensitive to reconstruction of electronic structure near $E_F$ as a result of the CDW formation.

\begin{figure*}[htb!]
    \includegraphics[width=0.88\textwidth]{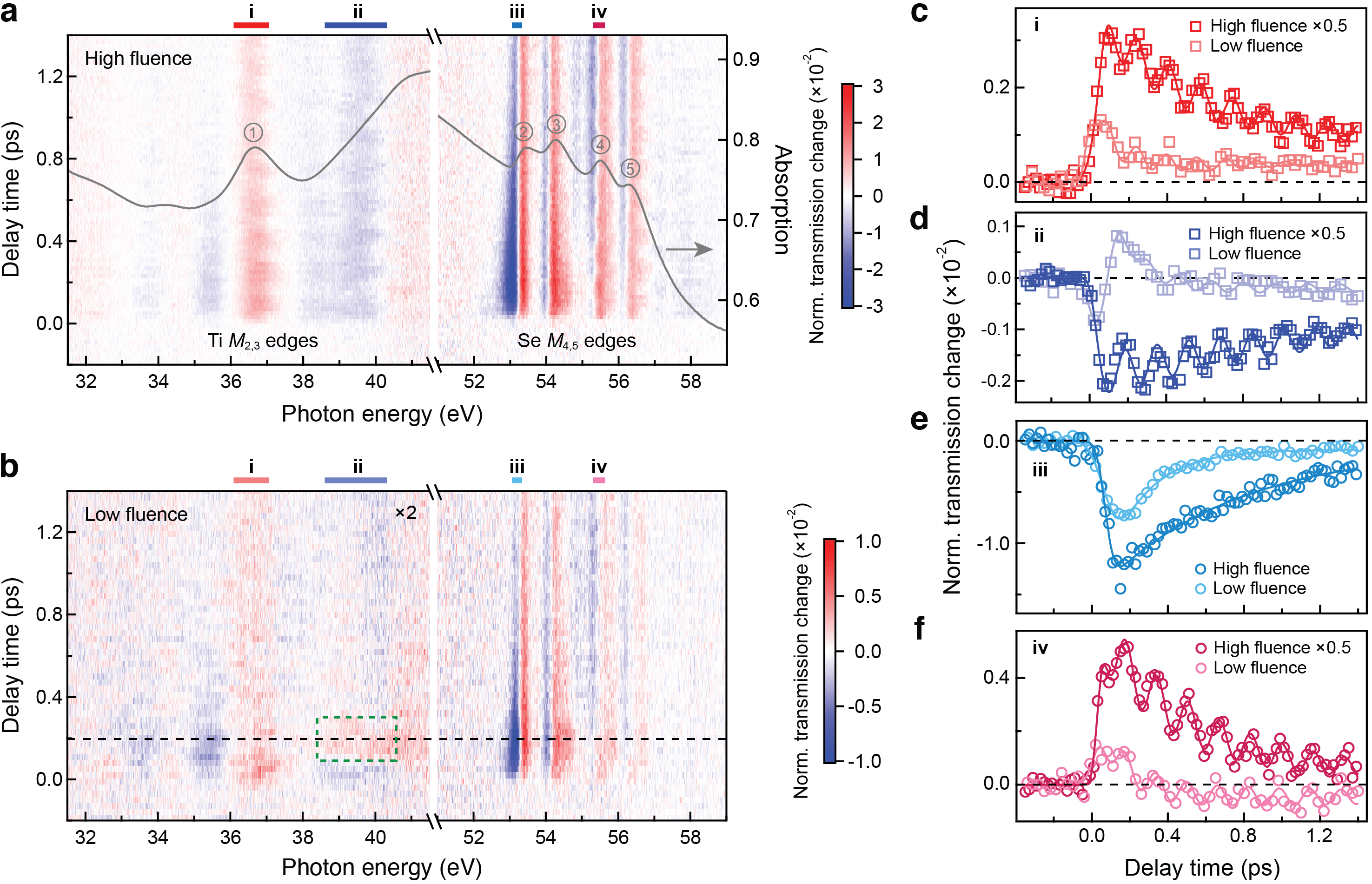}
    \caption{\textbf{Photoinduced dynamics of Ti and Se \textit{M} edges in the CDW phase.} \textbf{a},~Transient change in XUV transmission near the Ti~$M_{2,3}$ edges (left section) and Se~$M_{4,5}$ edges (right section) following photoexcitation by a 3.4-fs, 1.7-mJ/cm$^2$ pulse centered at 750~nm at a sample temperature of 23~K (see \ref{sn:xuv_setup} for experimental details). 
    The static absorption from Fig.~\ref{fig:1}j is reproduced in gray and plotted against the right axis. Horizontal bars on top indicate the four energy integration windows for the time traces in \textbf{c}--\textbf{f}; the spectral windows are \textbf{i}:~$[36.08,37.05]$~eV, \textbf{ii}:~$[38.60,40.32]$~eV, \textbf{iii}:~$[53.04,53.33]$~eV, and \textbf{iv}:~$[55.30,55.64]$~eV. \textbf{b},~The same as \textbf{a} but measured under a much reduced incident pump pulse fluence of 0.2~mJ/cm$^2$. Note the sign-flip in the transient signal highlighted by the green dashed box. The data for the Ti edges (left section) are multiplied by 2 to highlight this sign-flip. \textbf{c}--\textbf{f},~Transient XUV transmission dynamics in spectral windows~i--iv, measured for both low (lighter markers) and high (darker markers) incident fluences, where some data are multiplied by 0.5 for better visualization. Curves are fits to Eqs.~\eqref{eq:fit_erfexp} and \eqref{eq:fit_erfexp_cos}. Prominent 6.0~THz $A_{1g}$ coherent phonons are observed in all curves except in spectral window~iii of the Se~$M_5$ edge (panel~\textbf{e}).}
\label{fig:2}
\end{figure*}

\begin{figure*}[htb!]
    \includegraphics[scale=0.66]{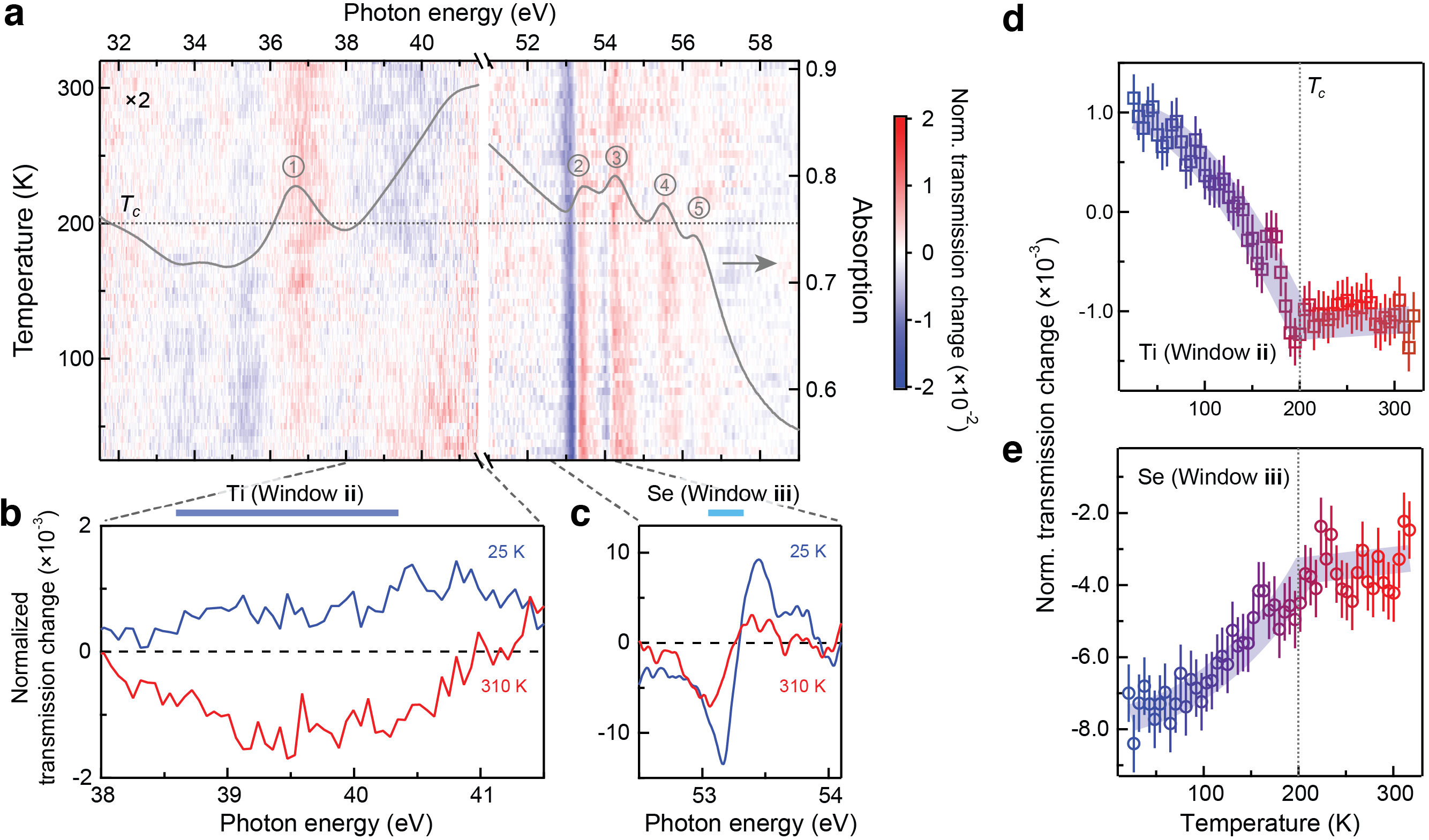}
    \caption{\textbf{Spectroscopic signatures of long-range order formation in time-resolved core-level absorption.} \textbf{a},~Temperature-dependent transient XUV transmission change at 0.2~ps (see dashed line in Fig.~\ref{fig:2}b). The incident fluence was 0.2~mJ/cm$^2$. The transient change was normalized by the transmission spectra before photoexcitation at each temperature measured. The data for the Ti edges (left section) is multiplied by 2. For reference, the static absorption from Fig.~\ref{fig:1}j is reproduced in gray and plotted against the right axis. \textbf{b},\textbf{c},~Selected transient XUV transmission change of Ti~$M_{2,3}$ edges (\textbf{b}) and Se~$M_5$ edge (\textbf{c}) both above (red) and below (blue) $T_c$. \textbf{d},\textbf{e},~Temperature-dependent transmission in selected spectral windows for Ti edges (window~ii in \textbf{b}) and Se edge (window~iii in \textbf{c}), both demonstrating a quasi-constant value above $T_c$ with an order-parameter-like onset below $T_c$. Error bars are 1~s.d. of values from repeated measurements at a given temperature. Blue curves are guides to the eye.}
\label{fig:3}
\end{figure*}

While the calculated density of states undergoes a visible change across the phase transition (see solid curves and filled shades in Fig.~\ref{fig:1}k), the core-level photoemission spectra stay almost constant across 300~K where we took a dense set of temperature points (Fig.~\ref{fig:static_core}a,c--e). The slight shifts of the Se~3$d$ edges by less than $30~$meV can be attributed to a global chemical potential change \cite{Chuang2020}, which is much less than the change in the CDW energy gap near the Fermi level ($\sim90$~meV, Fig.~\ref{fig:1}f) and crucially, it shows no spectroscopic singularity at $T_c$. This lack of change in the core-level photoemission spectra in 1$T$-TiSe$_2$ is in stark contrast to drastic changes seen in core-level spectra across other CDW transitions \cite{Wertheim1975,Hughes1976,Horiba2002,Han2023,Singh2018}. It was previously attributed to a lack of change in the local chemical environment in distinct sites of Ti and Se in the CDW state \cite{Negishi2006}, which is ascribed to a lack of inter-site charge transfer during the transition \cite{Chuang2020}. The absence of core-level photoemission changes at $T_c$ is echoed in the core-level absorption spectra for all five peaks identified (Fig.~\ref{fig:static_core}b,f--h). Aside from an overall temperature-dependent peak shift for the absorption edges \cite{Heinrich2023}, no distinct feature can be discerned at $T_c$, casting doubt on the sensitivity of core-level spectroscopies on detecting the low-energy phase transition in 1$T$-TiSe$_2$ that mostly concerns the electronic states near the Fermi level.

\section{S\lowercase{ensitivity of time-resolved core-level absorption to long-range order formation}}

The absence of equilibrium core-level features specific to the phase transition motivates us to search for its signature in the nonequilibrium state, where the added temporal dimension greatly expands the phase space so that we can examine core-level changes at selected times after the photoexcitation event. Figure~\ref{fig:2}a shows the representative XUV transmission change in the CDW state following photoexcitation by a sub-4-fs pulse with an incident fluence that far exceeds the known threshold for melting the periodic lattice distortion \cite{Porer2014,Hedayat2019,Duan2021,Burian2021,Cheng2022,Kurtz2024} (see \ref{sn:xuv_setup} for measurement details). Most transient features are found at or near the static absorption peaks \circled{1} to \circled{5}. Within the time window of our measurement, the most pronounced effect for all peaks is transient carrier excitation that leads to a larger population of electrons and holes above and below the Fermi level, respectively, which in turn results in enhanced and reduced XUV transmission at higher and lower photon energies along the rising edge of the static absorption peak. For the Se~$M_{4,5}$ edges, transient responses are similar between peaks~\circled{2} and \circled{3}, and between peaks~\circled{4} and \circled{5}. This similarity arises from the fact that each pair is related to the same valence states but different spin-orbit-split core-levels (see schematic in Fig.~\ref{fig:1}k). Besides the incoherent response from carrier excitation, coherent oscillatory responses with 6.0~THz frequency are observed across both Ti and Se edges, which are more clearly resolved in time traces (Fig.~\ref{fig:2}c--f) integrated over selected spectral windows~i--iv in Fig.~\ref{fig:2}a. The oscillation corresponds to the coherently excited $A_{1g}$ phonon where Se ions are displaced out-of-plane relative to a stationary Ti ion in the normal-state unit cell \cite{Holy1977}. Among the five absorption peaks \circled{1} to \circled{5}, only peaks~\circled{2} and \circled{3} do not exhibit the 6.0~THz oscillation (Fig.~\ref{fig:2}e). This lack of oscillation suggests a zero or minimal deformation potential between the $A_{1g}$ mode and the unoccupied Se~4$p$ bands near the Fermi level, making peaks~\circled{2} and \circled{3} ideal reporters of local electronic and possible excitonic dynamics free from the interference of coherent phonons.

Besides spectroscopic changes in peaks~\circled{1} to \circled{5}, the broad many-body continuum of the Ti edge also exhibits a clear photoinduced signal, where transient XUV transmission decreases (or increases) below (or above) 40~eV (see Fig.~\ref{fig:2}a). The broad nature of such changes that span more than 10~eV precludes a simple explanation based on carrier excitation and state filling, where the energy of the pump pulse restricts the excited carrier population to about 2~eV above the Fermi level. Instead, many-body dynamics such as photoinduced change in local screening is expected to influence the transient modification of XUV spectra \cite{Volkov2019,Schumacher2023}, making this energy window potentially sensitive to dynamics associated with excitonic correlations that accompany the CDW order. To test this hypothesis, we repeat the measurement by reducing the incident fluence nearly tenfold to 0.2~mJ/cm$^2$ so as to limit the number of photoexcited carriers that overwhelm the spectroscopic response at high fluences. While the most important effect of the fluence reduction is a near-proportional scaling-down of the transient changes (Fig.~\ref{fig:2}b--f), a qualitative difference is seen in the Ti many-body continuum peak. As highlighted by the time traces in Fig.~\ref{fig:2}d and the green dashed box in Fig.~\ref{fig:2}b, at 200~fs after photoexcitation, a sign reversal is observed in the transient XUV spectra, suggesting a separate many-body dynamical process that is distinct from simple carrier excitation.

To elicit the relation between this low-fluence feature in the Ti many-body continuum and the CDW phase transition, we fix the pump-probe delay at 200~fs (dashed line in Fig.~\ref{fig:2}b) with an incident fluence of 0.2~mJ/cm$^2$ and sweep the sample temperature in $\sim5$~K steps across $T_c$, as shown in Fig.~\ref{fig:3}a. Unlike the static core-level absorption spectra that show negligible changes across the phase transition (Fig.~\ref{fig:static_core}), the transient spectra are distinct above and below $T_c$ in this spectral window within the many-body continuum (Fig.~\ref{fig:3}b). Specifically, while the high-temperature XUV transmission change displays a transient decrease that is reminiscent of the high-fluence data in Fig.~\ref{fig:2}a, the transient XUV transmission increases at 25~K. Importantly, as shown in Fig.~\ref{fig:3}d, the temperature-dependent XUV transmission change in this spectral window exhibits an order-parameter-like onset similar to the valence-state observables such as the replica band spectral weight (Fig.~\ref{fig:1}e) and the CDW gap (Fig.~\ref{fig:1}f). This temperature-dependent onset unambiguously shows the sensitivity of the Ti many-body continuum to the establishment of long-range order, likely due to the integral role of valence charge excitations that produce this broad many-body peak, which are in turn modified by the changing Ti~3$d$ states near the Fermi level across the phase transition (Fig.~\ref{fig:1}k).

The core-level sensitivity to the phase transition is not restricted to the Ti edge and can also be discerned in the Se edge. Out of the four Se absorption peaks \circled{2}--\circled{5}, we focus on the transient transmission decrease in peak~\circled{2} (see spectral window~iii in Fig.~\ref{fig:2}a) because it is not affected by the $A_{1g}$ phonon and this feature is spectrally separated from other peaks. From the perspective of state-filling by excited carriers, this transmission decrease arises from the increased number of Se~4$p$ holes below the Fermi level that directly affect the excitonic channel of the phase transition, where Se~4$p$ holes are bound to Ti~3$d$ electrons. As photoexcitation is known to suppress the excitonic correlations \cite{Porer2014}, one can hence expect distinct transient responses of these Se~4$p$ holes above and below $T_c$. Indeed, comparing the transient spectral change of Se peak~\circled{2} at 310~K and 25~K (Fig.~\ref{fig:3}c), we see a clear difference in the amplitude of the intensity decrease. Specifically, a larger decrease occurs at 25~K compared to 310~K, and this distinction can be rationalized by the additional Se~4$p$ holes created via the breaking of condensed excitons in the ground state. From the temperature-dependent measurement (Fig.~\ref{fig:3}e), this changing amplitude of the transmission suppression shows a clear onset right at $T_c$, suggesting that the Se~$M_5$ edge is another core-level probe of long-range CDW formation due to its sensitivity to the transient population of Se~4$p$ holes.

\section{P\lowercase{ersistent excitonic fluctuations in the normal state}}

Having identified spectroscopic features in ABXAS about the establishment of long-range order, we leverage the intrinsic sensitivity of core-level spectroscopy to local charge density to detect possible presence of short-range order that serves as a precursor to the thermodynamic phase transition in this quasi-2D compound. Specifically, while mounting evidence such as thermal diffuse scattering in X-ray and electron diffractions \cite{Holt2001,Otto2021,Cheng2022,Cheng2024} has pointed out the presence of short-range lattice fluctuations, it remains to be seen whether short-range excitonic correlations are also present above $T_c$, which constitute a necessary condition for the excitonic character of the ground state in this quasi-2D system.

\begin{figure*}[htb!]
    \includegraphics[width=0.83\textwidth]{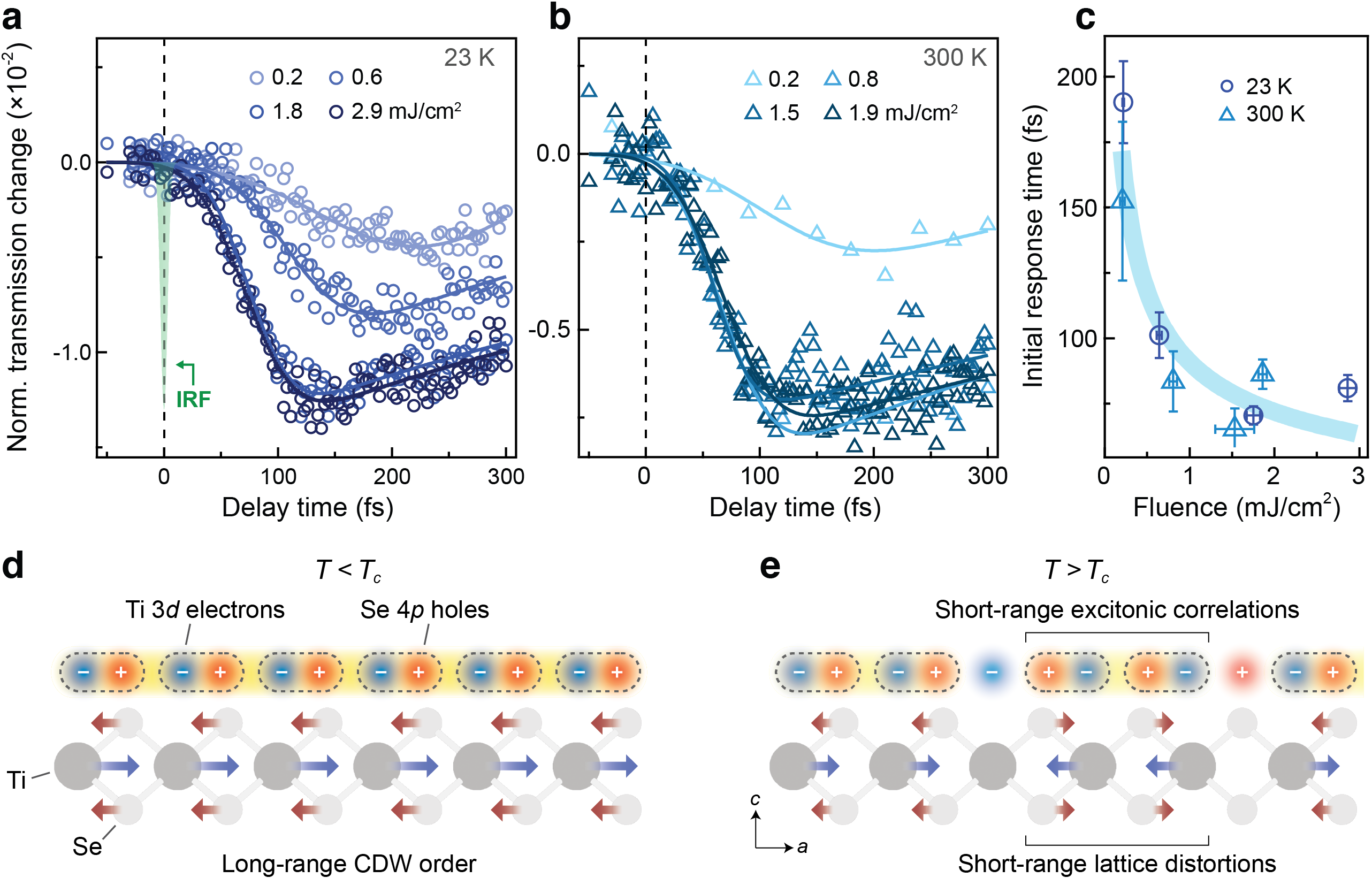}
    \caption{\textbf{Time-domain evidence for excitonic correlations.} \textbf{a},\textbf{b},~Short-time transient XUV transmission change for different pump fluences in the Se~$M_5$ edge, corresponding to spectral window~iii in Fig.~\ref{fig:2}a. Data in \textbf{a} were taken at 23~K in the CDW phase while that in \textbf{b} were measured at 300~K, which is above $T_c$. Curves are fits to Eq.~\eqref{eq:fit_erfexp}. The green shaded area in \textbf{a} represents the instrument response function (IRF) measured from helium gas transient (see Fig.~\ref{fig:IRF_He} 
    for details); it was vertically rescaled for visibility. \textbf{c},~Fluence-dependent initial response times $\tau$ extracted from \textbf{a} and \textbf{b}. Vertical error bars represent 1~s.d. of the fitting uncertainty, while horizontal error bars indicate the uncertainty of the pump fluence in use. Blue curve is a fit to $\tau \sim 1/\sqrt{F} + \tau_0$, where $F$ is the fluence and $\tau_0 = 26$~fs is a constant offset. \textbf{d},\textbf{e},~Schematic illustrations of excitonic correlations (\textit{top row}) and ionic displacements (\textit{bottom row}) in the CDW ground state (\textbf{d}) and the high-temperature state characterized by short-range fluctuations (\textbf{e}). Both short-range lattice distortions (represented by the arrows) and excitonic correlations (represented by the yellow-shaded dashed ovals) are present. For visual clarity, drawings in \textbf{d},\textbf{e} only focus on the effects of one particular CDW wavevector in a single layer out of the three equivalent wavevectors \cite{Cheng2024}.}
\label{fig:4}
\end{figure*}

Capitalizing on the high temporal resolution of ABXAS that is well below any material-intrinsic initial response time (see green curve in Fig.~\ref{fig:4}a and \ref{sn:xuv_setup}), we capture the time-domain signature of excitonic correlations by examining the characteristic time of the spectroscopic response and its dependence on the incident fluence. Here, the pump laser fluence effectively alters the mobile carrier density and hence the plasmon frequency, which in turn dictates the timescale of exciton breaking should local exciton correlations be present \cite{Rohwer2011,Hellmann2012,Okazaki2018}. To this end, a careful selection of the spectral window is needed to avoid the interference of the $A_{1g}$ phonon. For instance, in Ti peak~\circled{1} (window~i) and Se peak~\circled{4} (window~iv) where the responses are dominated by the phonon (Fig.~\ref{fig:Phonon_initial_response}a,b), the extracted initial response time is governed by the phonon frequency and hence independent from the pump laser fluence (Fig.~\ref{fig:Phonon_initial_response}c). Therefore, we again focus on Se peak~\circled{2} (window~iii) and probe how dynamics of the Se~4$p$ holes respond to varying carrier densities. The fluence-dependent result in the CDW state is shown in Fig.~\ref{fig:4}a, where the time taken for the initial suppression markedly decreases as the fluence increases. The fluence-dependent trend is consistent with previous time-resolved photoemission measurements \cite{Rohwer2011} that used a similar trend of timescales to illustrate the excitonic character of the ground state, as illustrated in Fig.~\ref{fig:4}d. The result thus reaffirms our earlier assignment of this spectral feature in Se peak~\circled{2} as a sensitive reporter of the excitonic contribution to the phase transition.

As Se peak~\circled{2} is dominated by local dipole transitions into the empty 4$p$ states, this time-domain signature of exciton breaking is expected to be agnostic about the existence of long-range phase coherence among the excitons. Hence, we can repeat these fluence-dependent measurements above $T_c$ to deduce possible presence of short-range excitonic correlations in the normal state, as illustrated in Fig.~\ref{fig:4}e. As shown in Fig.~\ref{fig:4}b, at 300~K, the initial response visibly slows down with a lower incident fluence, reproducing the trend measured below $T_c$ within experimental uncertainties. For datasets taken both above and below $T_c$, the fluence-dependent initial response times follow the $1/\sqrt{F}$ scaling with a constant vertical offset (blue curve in Fig.~\ref{fig:4}c), where $F$ denotes the incident fluence and is proportional to the excited mobile carrier density $n$ (ref.~\cite{Rohwer2011}). As the plasmon frequency follows $\sqrt{n}$, this $1/\sqrt{F}$ scaling suggests the presence of excitonic correlations in both the CDW and normal states, whereas the constant offset can be attributed to time needed for creating mobile Se~4$p$ holes in the absence of excitonic interactions.

\section{D\lowercase{iscussion}}

Our detection of short-range excitonic correlation seen from the time-domain response in core-level absorption provides important evidence to resolve the debate in equilibrium experiments on whether such correlations are present. Momentum-resolved electron energy-loss spectroscopy first reported signatures of exciton condensate by identifying a plasmon mode as its collective excitation, which softens at $T_c$ (ref.~\cite{Kogar2017}). A more recent study using a similar technique, however, reported no such softening behavior, where Landau damping at a finite momentum prevents a clear resolution of this plasmon mode \cite{Lin2022}. While the excitonic character of the ground state is supported by previous time-resolved photoemission measurements that monitored the time taken for the photoinduced suppression of the replica Se bands \cite{Rohwer2011}, this approach does not work above $T_c$ due to the very weak intensity and incoherent nature of the replica bands in the absence of long-range order (Fig.~\ref{fig:1}c). In this regard, the advantage of ABXAS lies in its intrinsic sensitivity to local excitonic correlations, which are present in both short-range and long-range ordered states. 

Given the strong coupling between excitonic and phononic degrees of freedom in this compound \cite{VanWezel2010,Porer2014,Kogar2017,Hedayat2019,Lian2020}, there remains the possibility that the fluence-dependent initial response times seen in Fig.~\ref{fig:4}a--c can be caused by CDW-related lattice distortions in addition to being affected by exciton breaking. In this scenario, as the fluence increases, transient hardening of the CDW phonon can potentially result in a perceived shortening of the initial response time, making it difficult to distinguish short-range excitonic correlations from short-range lattice fluctuations (Fig.~\ref{fig:4}e), the latter of which are known to be present above $T_c$ (refs.~\cite{Holt2001,Otto2021,Cheng2022,Cheng2024}). However, we can confidently exclude any dominant phonon contribution to the decreasing response time seen in Fig.~\ref{fig:4}c because transient renormalization of the CDW phonon frequency in our photoexcitation regime was measured to be 20\% (ref.~\cite{Otto2021}), corresponding to a 20\% decrease in the initial response time. This change is dwarfed by more than threefold reduction of the observed initial response time (see blue curve in Fig.~\ref{fig:4}c), suggesting that CDW phonon contributions, if any, cannot be the leading factor.

Due to the quasi-2D nature of 1$T$-TiSe$_2$, we expect the observed excitonic correlations in the normal state to be mostly confined in-plane, contributing to the short-range CDW fluctuations that lack 3D phase coherence. From previous fluence-dependent ultrafast electron scattering study \cite{Cheng2022}, it was further shown that exciton correlations are responsible for maintaining the inter-plane phase coherence of the CDW ground state. Given the new insight from ABXAS, we thus conclude that exciton correlations are instrumental in both the amplitude and phase coherence of the CDW state.

Our work offers the first demonstration where an attosecond core-level technique traditionally applied to the study of high-energy dynamics can simultaneously probe low-energy phase transitions in quantum materials. The temporal resolution allows the identification of spectroscopic features in the time domain that are sensitive to the establishment of long-range order whereas the corresponding equilibrium measurements fail to capture any spectroscopic singularity. The strength of our probe also lies in its intrinsic sensitivity to local electronic environments, helping us pinpoint the elusive observable of excitonic correlations in the normal state of 1$T$-TiSe$_2$, which hold the key to understand exciton condensation in a reduced dimension. With access by ABXAS to different energy-, time-, and length-scales, we envision attosecond core-level spectroscopies to play a crucial role in the investigation of strongly-correlated materials where multi-scale dynamics and interactions are hallmark features to be understood.\\


%

\section{A\lowercase{dditional information}}

\noindent\textbf{Acknowledgements.}~
We thank Sasawat~Jamnuch, Tod~A.~Pascal, Luca~Moreschini, and Anshul~Kogar for helpful discussions, and Donghui~Lu and Makoto~Hashimoto for assistance in the photoemission measurements at Beamline~5-2 at SSRL. We thank Holger~Oertel, William~Windsor, Laurenz~Rettig, Ralph~Ernstorfer, and Martin~Wolf from Fritz Haber Institute  (FHI) Berlin and technical staff at FHI for support and discussion that enabled the construction of the cryogenic endstation in the ABXAS instrument.
A.Z. acknowledges support from the Miller Institute for Basic Research in Science and the National Science Foundation (NSF-DMR~2247363).
S.-C.L. acknowledges support by the Berkeley-Taiwan Fellowship and the National Science Foundation (NSF-DMR 2247363).
This work utilized the computational resources of the HPC systems at the Max Planck Computing and Data Facility (MPCDF) and the Supercomputer Center at the Institute for Solid State Physics, University of Tokyo. 
The work at SSRL is supported by the U.S. Department of Energy (DOE), Office of Science, Office of Basic Energy Sciences, Division of Materials Sciences and Engineering under contract DE-AC02-76SF00515.
B.L. acknowledges support from the Ministry of Science and Technology of China (2023YFA1407400) and the National Natural Science Foundation of China (12374063). 
D.X. acknowledges support from the National Natural Science Foundation of China (Nos.~11925505, 12335010) and the New Cornerstone Science Foundation through the XPLORER PRIZE.
W.X. and Y.G. acknowledge the Shanghai Science and Technology Innovation Action Plan (Grant No.~21JC1402000) and the Double First-Class Initiative Fund of ShanghaiTech University.
E.B. and B.R.N. acknowledge support from the National Science Foundation Graduate Research Fellowships Program under Grant Nos.~DGE~2146752 and 1752814, respectively.
M.H. acknowledges funding by the National Science Foundation (NSF-REU EEC-1852537). 
Any opinions, findings, and conclusions or recommendations expressed in this material are those of the authors and do not necessarily reflect the views of the National Science Foundation. 
M.W.Z. acknowledges funding by the W.~M. Keck Foundation, Laboratory Directed Research and Development Program at Berkeley Lab (107573 and 108232), the Max Planck Society, and the Hellman Fellows Fund which enabled the building of the ABXAS instrument. M.W.Z. further acknowledges support by the Department of Energy (DE-SC0024123).\\

\noindent\textbf{Author contributions.}~
A.Z. and M.W.Z. conceived the project. A.Z., S.-C.L., E.B., B.R.N., M.H., and M.W.Z. designed and constructed the attosecond broadband extreme-ultraviolet absorption spectroscopy beamline. A.Z. and S.-C.L. performed the static and time-resolved core-level absorption spectroscopy experiments. A.Z. and B.Q.L. performed the angle-resolved photoemission experiments. A.Z. and S.-C.L. analyzed the data. S.A.S. performed static and time-dependent density functional theory calculations. Y.C., W.X., Y.G., and D.X. prepared the samples used in this study. A.Z., S.-C.L., and M.W.Z. wrote the manuscript with inputs from all co-authors. The research was supervised by M.W.Z.\\

\noindent\textbf{Competing interests.}~
The authors declare no competing interests. \\

\noindent\textbf{Data availability.}~
All of the data supporting the conclusions are available within the article and the Supplementary Information. Additional data are available from the corresponding author upon reasonable request.

\clearpage
\newpage

\beginsupplement

\onecolumngrid
\begin{center}
\textbf{\large Supplementary Information for ``\papertitle''}
\end{center}

\vspace{0.8cm}
\twocolumngrid

\section{S\lowercase{ample growth and preparation\label{sn:sample}}}

High-quality single crystals of 1$T$-TiSe$_2$ were grown by chemical vapour transport with an iodine transport agent. Ti and Se were mixed in a molar ratio of 1:2 and placed into an alumina crucible before being sealed into a quartz tube. The quartz tube was heated to 700$^\circ$C and 1$T$-TiSe$_2$ crystals were synthesized at the 650$^\circ$C zone for two weeks. For absorption spectroscopy measurements, 1$T$-TiSe$_2$ thin flakes were obtained by repeated exfoliation of the bulk crystal with polydimethylsiloxane films (PDMS, Gel-Pak). Flakes were pre-screened for thickness and uniformity with an optical microscope using the color contrast and further characterized by atomic force microscopy. Selected flakes were detached from PDMS in ethanol and scooped onto copper TEM grids. The resulting free-standing flake has a typical lateral dimension of approximately $400~\upmu$m and thickness around 30~nm. An optical image of a measured flake is shown in Fig.~\ref{fig:1}h.

\section{A\lowercase{ngle-resolved photoemission measurements\label{sn:arpes}}}

\begin{figure}[htb!]
    \includegraphics[width=\columnwidth]{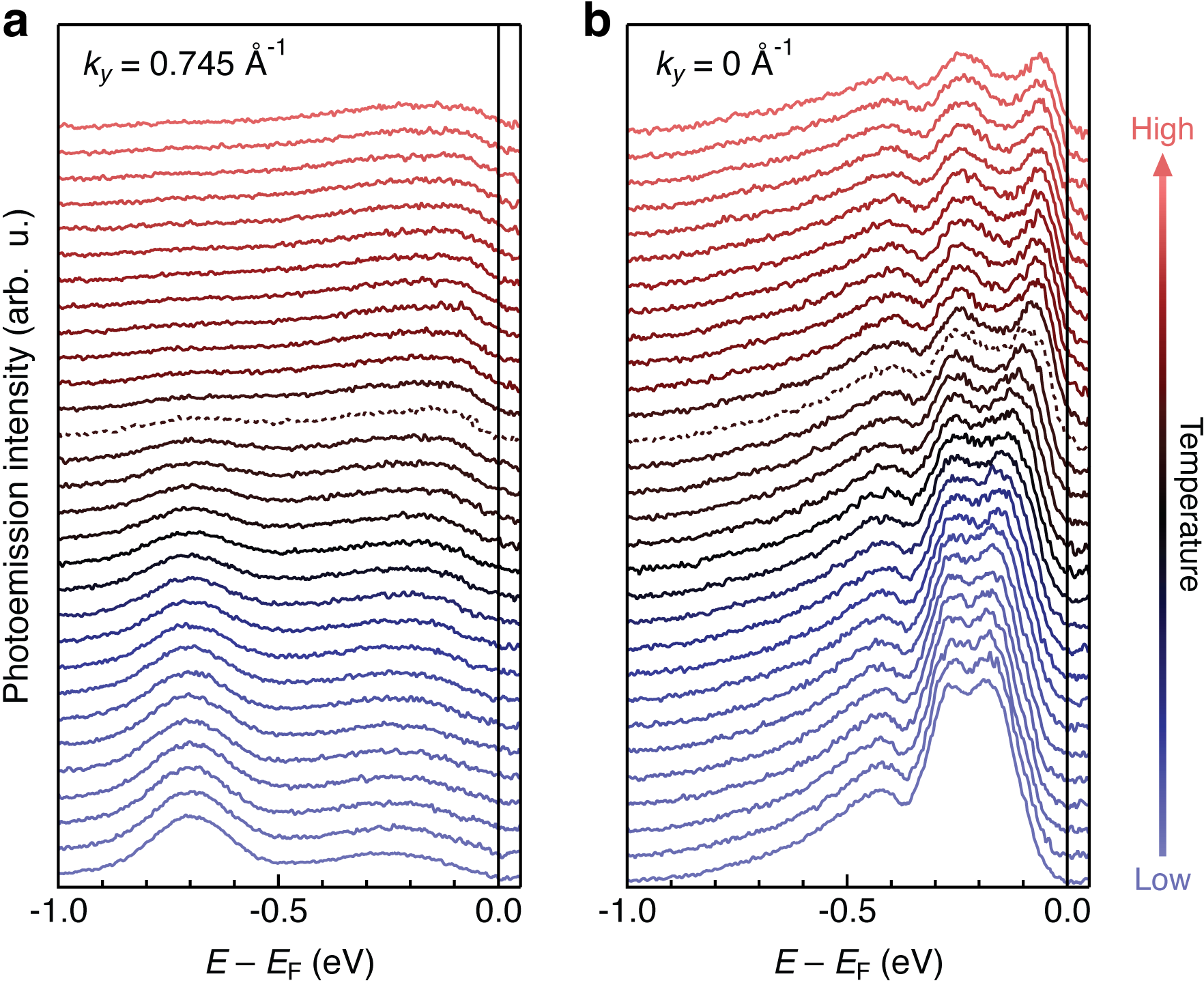}
    \caption{\textbf{Change of valence electronic structures across the phase transition.} \textbf{a},~Temperature-dependent energy distribution curves (EDCs) taken at $k_y = 0.745~${\AA}$^{-1}$ where the appearance of coherent replica Se~4$p$ bands manifests as a hump around $-0.7$~eV at low temperature (see Fig.~\ref{fig:1}d). Photoemission intensity within a 0.05~{\AA}$^{-1}$ window was integrated for each curve. \textbf{b},~Temperature-dependent EDCs at the $\Gamma$ point, showing the gap opening that manifests as a shift of the leading edge to higher binding energies when temperature decreases. Photoemission intensity within a 0.025~{\AA}$^{-1}$ window was integrated for each curve. In both \textbf{a} and \textbf{b}, EDCs are vertically offset for clarity and both vertical axes use a linear scale. Each EDC corresponds to a data point in the temperature-dependent plots in Fig.~\ref{fig:1}e,f. The EDC near $T_c \approx 200$~K is shown as a dashed curve in each panel.}
\label{fig:static_valence}
\end{figure}

High-resolution angle-resolved photoemission spectroscopy was performed at Beamline~5-2 of the Stanford Synchrotron Radiation Lightsource (SSRL) at SLAC National Accelerator Laboratory with a Scienta Omicron DA30L electron analyzer. The photon energy was 95~eV, and the combined beamline and analyzer energy resolution was below 22~meV in all measurements. The photon beam spot had an approximate cross-sectional dimension smaller than $50~\upmu\text{m}\times30~\upmu\text{m}$. The chemical potential was independently determined from the spectra of a polycrystalline gold that was electrically connected to the sample. Fresh (001) surfaces of 1$T$-TiSe$_2$ were obtained by cleaving the crystal \textit{in~situ} in ultrahigh vacuum. The pressure was maintained below $3.5\times10^{-11}$~Torr for all temperatures investigated thanks to a local heating device placed in close proximity to the sample so that most parts of the sample manipulator and radiation shield were maintained at a substantially lower temperature compared to the sample, acting as a cold trap. Multiple temperature cycles were performed to verify negligible sample surface degradation and to reproduce the observed temperature-dependent trends in Fig.~\ref{fig:1}. All photoemission data presented were taken with a linear horizontal polarization (LH, within the photoemission plane). All constant-energy maps shown have an intensity integration window of 5~meV centered at the specified energy. 

\begin{figure*}[htb!]
    \includegraphics[width=0.80\textwidth]{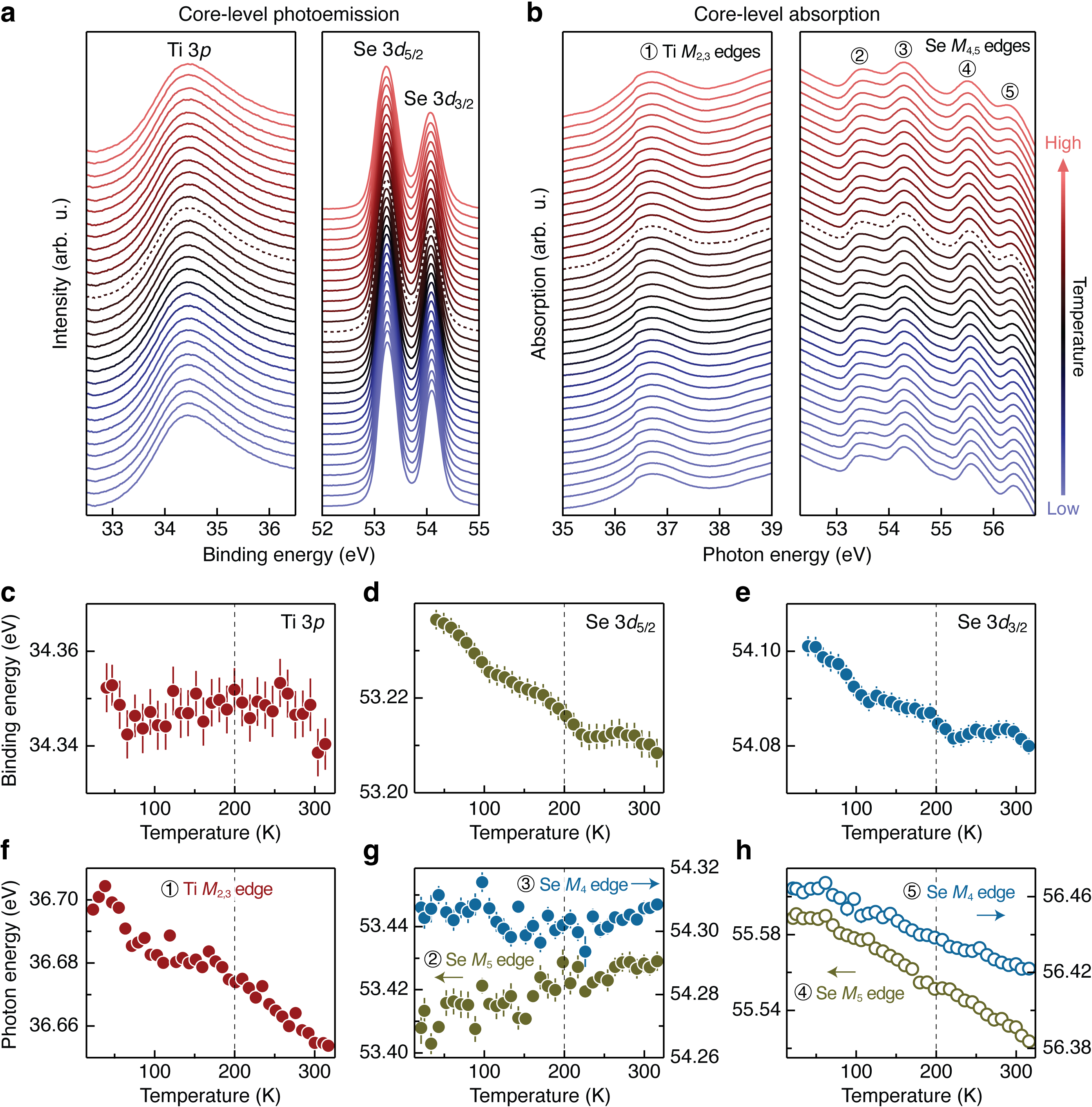}
    \caption{\textbf{Absence of phase transition signals in equilibrium core-level spectroscopies.} \textbf{a},\textbf{b},~Temperature-dependent photoemission intensity for the Ti~3$p$ core level (\textit{left}) and spin-orbit-split Se~3$d$ core levels (\textit{right}). \textbf{c},\textbf{d},~Temperature-dependent XUV absorption for the Ti~$M_{2,3}$ edges (\textit{left}) and Se~$M_{4,5}$ edges (\textit{right}). In panels~\textbf{a} and \textbf{b}, each curve is vertically offset for clarity and all vertical axes use a linear scale. Each curve corresponds to a data point in the corresponding quantity in panels~\textbf{c}--\textbf{h}. The curve near $T_c \approx 200$~K is shown as a dashed curve in each panel. \textbf{c}--\textbf{h},~Temperature-dependent photoemission and absorption peak positions extracted from \textbf{a} and \textbf{b}. Aside from a global temperature-dependent peak shift in certain core-level and absorption peaks, there is no observable spectroscopic singularity at $T_c$. Error bars represent 1~s.d. of the fitting uncertainty for determining the peak positions.}
\label{fig:static_core}
\end{figure*}

To determine the spectral weight of the folded Se~4$p$ bands, we integrate over the photoemission intensity in the dashed box in Fig.~\ref{fig:1}d for all temperatures. In practice, we first extracted the energy distribution curves (EDCs) at the corresponding momentum (Fig.~\ref{fig:static_valence}a) and then summed over the intensity in the energy window $[-0.866,-0.553]$~eV relative to the Fermi level. To determine the energy gap $E_\text{gap}$ in Fig.~\ref{fig:1}f, the leading edge position of the EDCs at the $\Gamma$ point, shown in Fig.~\ref{fig:static_valence}b, was computed from the minimum position of their first energy derivatives. For all core-level photoemission spectra presented in this work, they were obtained from angle-resolved data via momentum integration over a window of $\pm0.75$~{\AA}$^{-1}$, and the peaks were determined by fitting to a Lorentzian function with a linear background. Temperature-dependent core photoemission spectra are shown in Fig.~\ref{fig:static_core}a,c--e. As mentioned in the main text, no evidence for the long-range order formation at $T_c$ can be discerned from these core spectra.

\section{B\lowercase{roadband} XUV \lowercase{absorption spectroscopy experiments}\label{sn:xuv_setup}}

\subsection{Broadband XUV generation}

\begin{figure*}[tbh!]
	\includegraphics[scale=0.62]{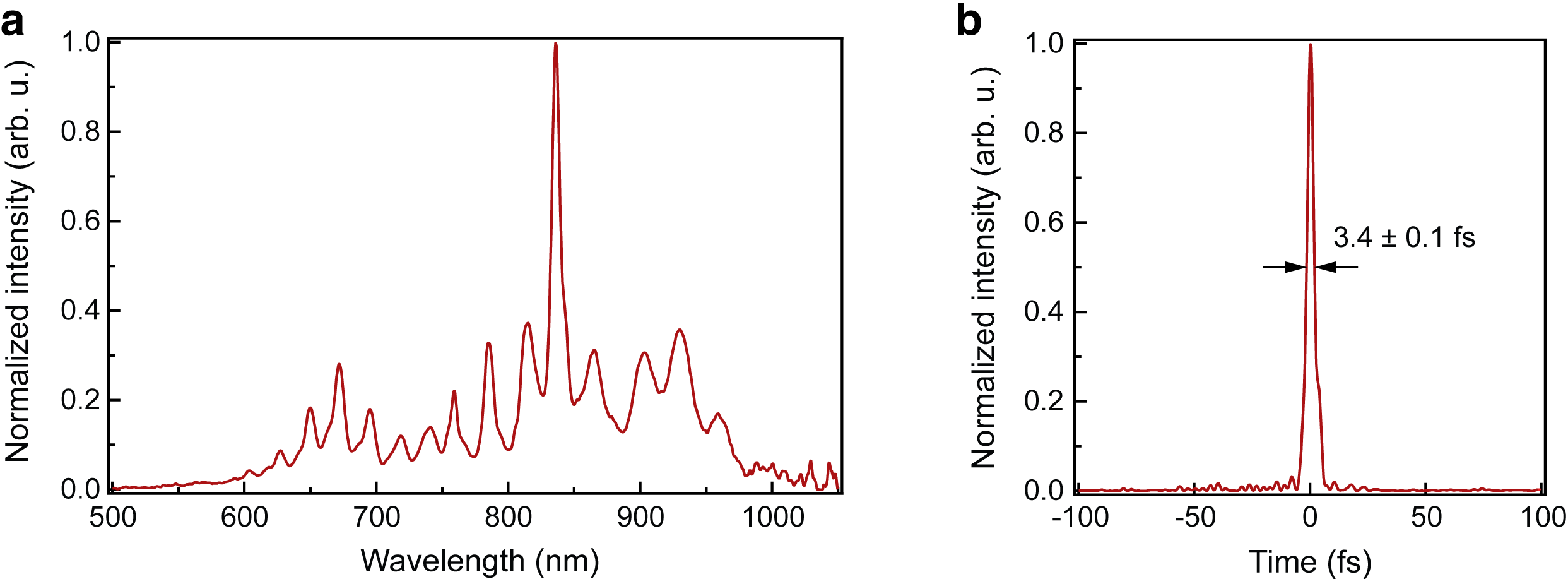}
	\caption{Normalized spectrum (\textbf{a}) and retrieved temporal profile (\textbf{b}) of the compressed NIR pulse after the He-filled hollow-core fiber and the chirped mirror array. Pulse characterization was performed using dispersion scans \cite{Silva2014}.}
\label{fig:NIR_pulse_profile}
\end{figure*}

\begin{figure*}[tbh!]
    \includegraphics[scale=0.62]{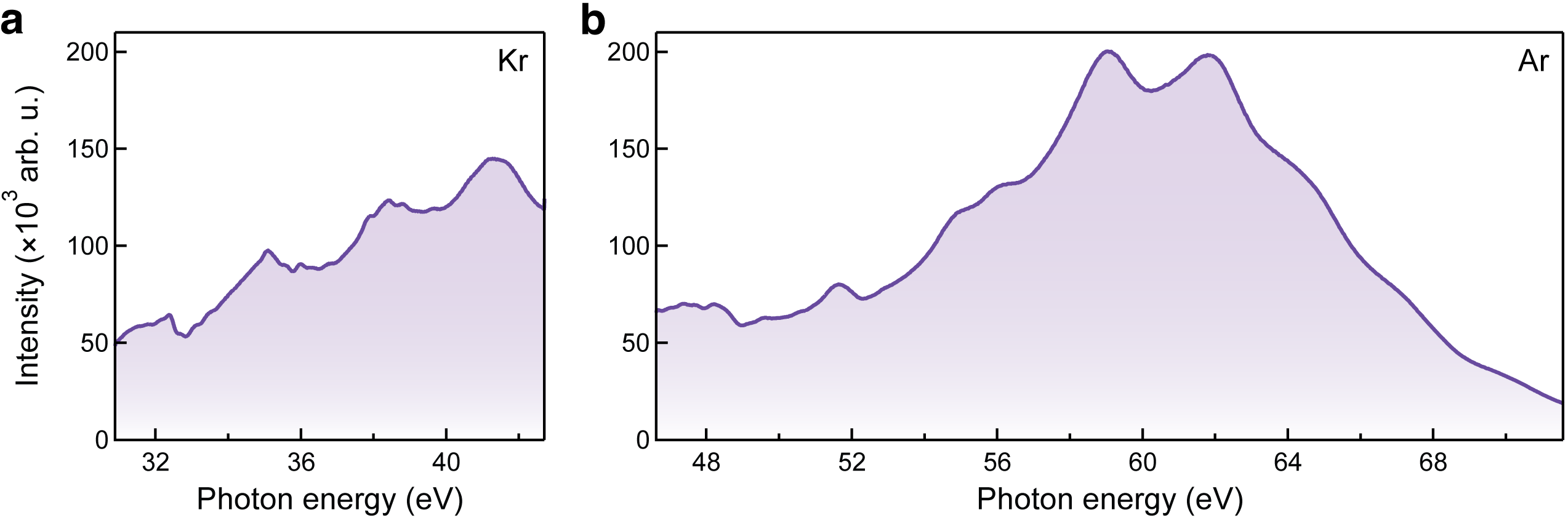}
    \caption{Typical XUV source spectra obtained from high harmonic generation in the in-house ABXAS beamline using either Kr (\textbf{a}) or Ar (\textbf{b}) with a gas-cell pressure around 40--60~Torr. The typical incident NIR pulse energy was 1.5--2~mJ. The spectra were optimized for maximum counts near the Ti and Se~$M$ edges under investigation.}
\label{fig:HHG_source}
\end{figure*}

The static and transient XUV measurements were conducted using a table-top setup at UC Berkeley. To generate the XUV light, 5~mJ near-infrared (NIR) pulses centered at 790~nm from a Ti:sapphire amplifier system operating at 1~kHz repetition rate (Legend Elite Duo HE, Coherent Inc.) were focused into a stretched hollow-core fiber (700~$\upmu$m inner diameter and 6.4~m length), which was filled with 20~psi of ultrahigh purity He for spectral broadening. The resulting broadband pulses, spanning from approximately 500 to 1000~nm and with a typical pulse energy of 3~mJ, were temporally compressed by an 8-pair chirped mirror array (PC1332, Ultrafast Innovations) and a wedge pair (d-scan, Sphere Ultrafast Photonics). The resulting NIR pulse has a typical pulse duration of $3.4\pm0.1$~fs (measured at FWHM), as characterized by the dispersion scan \cite{Silva2014}. The compressed pulse spectrum and temporal profile are shown in Fig.~\ref{fig:NIR_pulse_profile}.

The compressed NIR beam was subsequently steered through a thin fused-silica window into a vacuum chamber and then divided into probe and pump arms with a 90:10 intensity ratio using a custom antireflection-coated beamsplitter (Layertec GmbH). The probe was focused onto a gas cell at which either Ar or Kr was applied to generate broadband XUV light via high harmonic generation (HHG). Typical HHG spectra from the Kr and Ar source are shown in Fig.~\ref{fig:HHG_source}a,b, respectively, demonstrating the broad bandwidth that corresponds to one or few attosecond XUV pulses in the time domain \cite{Krausz2009}. The resulting XUV pulses traveled through a 0.2~$\upmu$m thick Al foil that removed residual NIR light. Using an Ni-coated toroidal mirror at grazing incidence and in a $2f$:$2f$ configuration, the XUV pulses were subsequently refocused into a sample chamber, in which a motorized three-axis stage was used to translate the samples. The transmitted XUV beam through the sample was steered towards a flat field concave grating (Hitachi 001-0660) and dispersed onto an XUV charge-coupled device (ALEX, Greateyes GmbH).

\subsection{Static absorption measurement}

To obtain static absorption spectra like those shown in Figs.~\ref{fig:1}j and \ref{fig:static_core}b, the transmitted XUV intensity after the sample as a function of photon energy [$I_\text{sample}(E)$] was referenced to that of a blank region [$I_\text{blank}(E)$] on the same copper TEM grids where the sample was placed; see Fig.~\ref{fig:1}h for an optical micrograph of the mounted sample.
With this reference, XUV absorption by the sample is calculated as $1 - I_\text{sample}(E)/I_\text{blank}(E)$, where we leverage the close-to-zero reflectivity of XUV light at normal incidence. 

\begin{figure*}[tbh!]
	\includegraphics[scale=0.69]{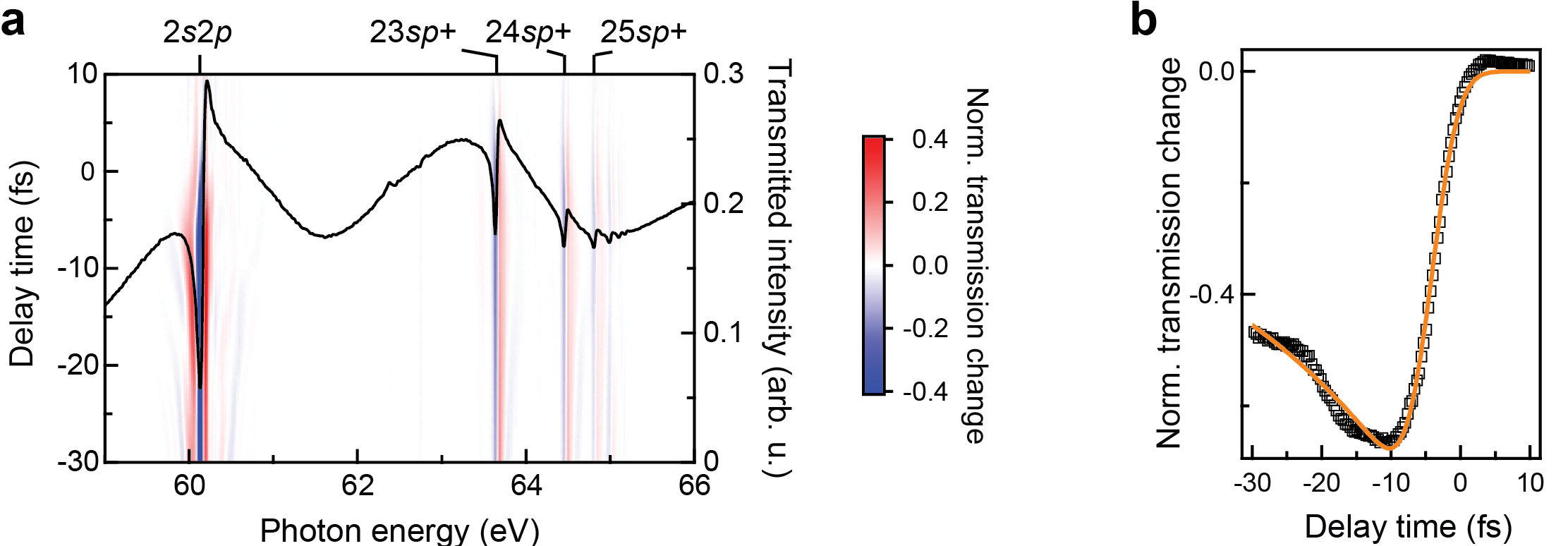}
    \caption{\textbf{Characterization of ABXAS temporal resolution.} \textbf{a},~Transient transmission spectrogram of He two-electron excitation states. The equilibrium transmitted XUV spectrum was plotted against the right axis as a reference for the corresponding states, which are labeled at the top. Positive (negative) time delay means that XUV pulse arrives after (before) the NIR pump pulse. This measurement used 30~Torr of He, measured before it exited the gas cell. \textbf{b},~Transient signal of the He $2s2p$ state extracted from \textbf{a}. Yellow curve is a fit to Eq.~\eqref{eq:fit_erfexp}.}
\label{fig:IRF_He}
\end{figure*}

\subsection{Time-resolved broadband XUV absorption}

For the ABXAS measurements, an in-vacuum piezo-controlled delay stage (CLL42, Smaract Inc.) was employed to control the temporal delay between the pump and probe pulses. After the delay stage, the pump traveled through a mechanical shutter for rapid acquisition of spectra with and without the pump. A motorized iris (SID5714, SmarAct Inc.) was used to finely adjust the pump fluence, which was accurately determined by measuring the pulse energy (by EnergyMax-RS J-10MB-LE, Coherent Inc.) and spatial profile (by S-WCD-UHR, DataRay) of an image beam. The pump beam spot size on the sample was varied between 400 and 700~$\upmu$m (FWHM) due to the use of the motorized iris for fluence adjustment; in all cases, this size was much larger than the XUV probe beam size (approximately 70~$\upmu$m at FWHM) to ensure near-uniform excitation in the probed volume. To filter out residual pump light scattered off the sample, another 0.2~$\upmu$m thick Al foil was installed between the sample chamber and the XUV spectrometer. Spatial overlap between the pump and probe beams was determined through an on-target pinhole, which was imaged by a sample-monitoring camera equipped with a long working-distance objective.

\subsection{Characterization of temporal resolution of the ABXAS beamline}

Temporal overlap between the NIR pump and XUV probe as well as the instrument response function (IRF) were obtained by measuring the transient absorption changes in He. As shown in Fig.~\ref{fig:IRF_He}a, the absorption profile of the doubly excited states in He can be dramatically changed by an intense NIR beam; the corresponding transmitted spectrum in equilibrium without NIR excitation is also shown in Fig.~\ref{fig:IRF_He}a as the black curve \cite{Madden1963}. Upon NIR photoexcitation, the Autler-Townes splitting of these excited states was clearly observed as a function of delay time. By taking a line cut at the 2$s$2$p$ state at 60.11~eV, the temporal profile of the state dressed by the NIR photons was shown in Fig.~\ref{fig:IRF_He}b. The asymmetric temporal profile was fit to Eq.~\eqref{eq:fit_erfexp}, and the width of the error function $\omega = 7.2\pm0.2$~fs was quoted in the main text as the temporal resolution \cite{Zurch2017}, where we assumed that the timescale of the initial response in the absorption change is limited by the instrument response.

\subsection{Temperature-dependent measurements}
A vibration-isolated closed-cycle liquid-He cryostat system (SHI-4XG-UHV-2, Coldedge Inc.) was used to control and maintain the sample temperature along with a Lake Shore Cryotronics controller for closed PID loop operation. To avoid condensation, a base pressure of $1.1 \times 10^{-9}$~Torr was achieved in the sample chamber. An oxygen-free high thermal conductivity (OFHC) copper sample holder was connected to the cold finger of the cryostat through an OFHC copper braid. The effective minimum temperature reached on-target in our system was around 20~K, providing a wide temperature window from 320~K to 20~K to study 1$T$-TiSe$_2$ whose $T_\text{c}$ is around 200~K. We note that for fidelity reasons no special attention was given to radiation shielding in this experiment, preventing sample temperature from reaching the cryostat base temperature of sub-10~K. To ensure that the same part of the sample was photoexcited and probed at all temperatures, four piezo-controlled mirrors paired with quad-diodes (MRC Systems, GmbH) were installed to ensure beam pointing stability, and an automatic sample drift correction system was implemented to counter thermal expansion and shrinkage. The drift correction system is based on an efficient sub-pixel image registration algorithm \cite{Guizar2008} that works together with the sample-monitoring camera and the three-axis motorized sample stage.

\begin{figure*}[tbh!]
    \includegraphics[scale=0.66]{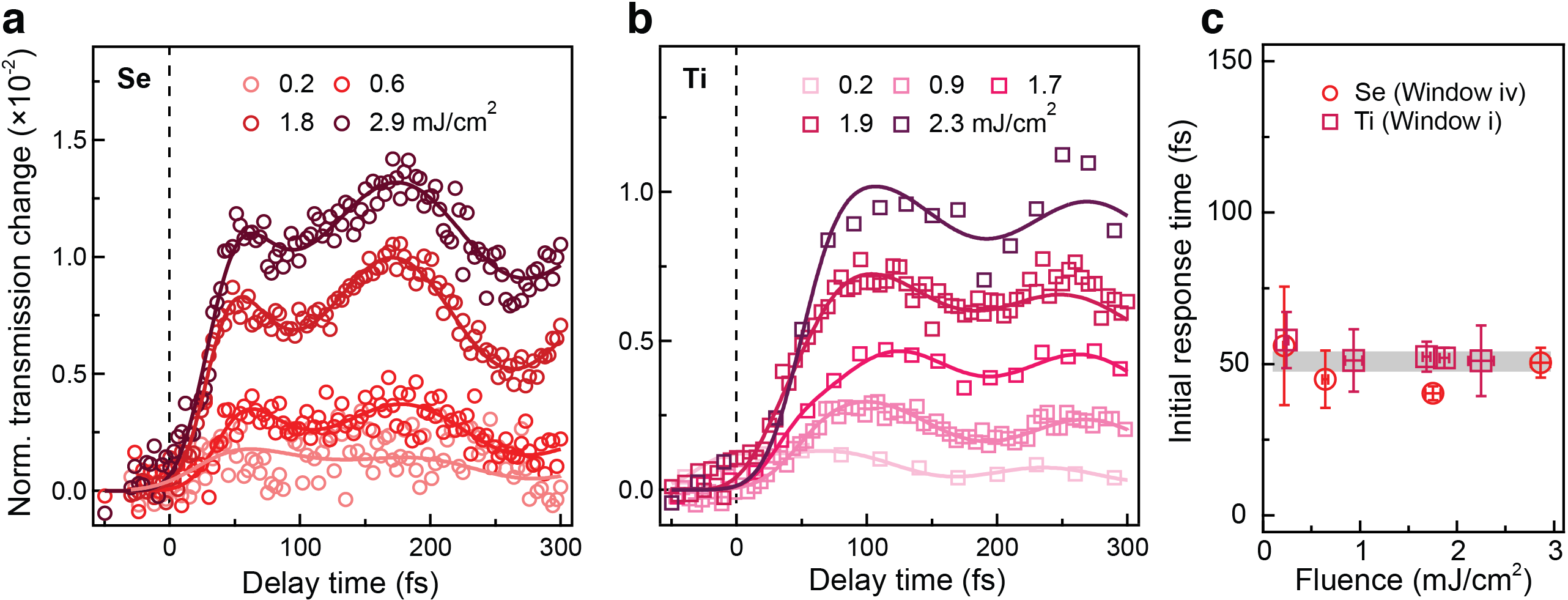}
    \caption{\textbf{Oscillatory absorption transients dominated by the coherent \textit{A}$_\text{1\textit{g}}$ phonon.} \textbf{a},\textbf{b},~Short-time transient XUV transmission change for different pump fluences in Se (\textbf{a}, window~iv) and Ti (\textbf{b}, window~i) $M$ edges. The oscillatory features originate from the $A_{1g}$ phonon, and solid curves are fits to Eq.~\eqref{eq:fit_erfexp_cos}. \textbf{c},~Fluence-dependent initial response times, $\omega$ in Eq.~\eqref{eq:fit_erfexp_cos}, extracted from \textbf{a} and \textbf{b}. Vertical error bars represent 1 s.d. of the fitting uncertainty, and horizontal error bars indicate the uncertainty of the pump fluence in use. The initial response time does not vary with fluence within the measurement error, as highlighted by the gray horizontal line that represents the average of all data points.}
\label{fig:Phonon_initial_response}
\end{figure*}

\subsection{XUV spectrometer calibration}

Using a gas cell integrated into the sample holder assembly, we calibrated the energy in the static absorption spectra by known energies of autoionizing levels in He, Ne, and Ar, as well as by the Al $L_\text{2,3}$ edges created by the NIR filter. The 13 autoionizing states of the noble gases used for calibration are listed in Table~\ref{table:calibration}, covering the entire spectral range of interest for Ti and Se~$M$ edges studied here.

\begin{table}[tbp!]
\centering
\caption{Selected atomic levels used for the calibration of the XUV spectrometer.}
\label{table:calibration}
\begin{footnotesize}
\begin{tabular}{>{\centering\arraybackslash}p{1.5cm} >{\centering\arraybackslash}p{2.5cm} >{\centering\arraybackslash}p{2.5cm} >{\centering\arraybackslash}p{0.5cm}}
\toprule
\textbf{Gas} & \textbf{States} & \textbf{Energy (eV)} & \textbf{Ref.} \\
\midrule
\addlinespace[0.5em]
\multirow{3}{*}{\centering He} & $24sp+$ & 64.465(7) & \multirow{3}{*}{\citenum{Lipsky1966}} \\
                         & $23sp+$ & 63.654(6) & \\
                         & $2s2p$   & 60.126(15) & \\
\midrule
\addlinespace[0.5em]
\multirow{5}{*}{\centering Ne} & $2s^12p^66p$ & 47.9650(30) & \multirow{5}{*}{\citenum{Schulz1996}} \\
                         & $2s^12p^65p$ & 47.6952(15) & \\
                         & $2s^12p^64p$ & 47.1193(50) & \\
                         & $2s^12p^63p$ & 45.5442(50) & \\
                         & $2p^43s3p$   & 44.9817(50) & \\
\midrule
\addlinespace[0.5em]
\multirow{5}{*}{\centering Ar} & $(^1S)3d(^2D_{5/2})4p$ & 34.994(2)  & \multirow{5}{*}{\citenum{Madden1969}} \\
                         & $(^1D)3d(^2P_{3/2})4p$ & 34.412(4) & \\
                         & $(^3P)3d(^2D_{3/2})4p$ & 31.624(2)  & \\
                         & $(^3P)4s(^2P_{1/2})5p$ & 31.602(2)  & \\
                         & $(^1D)4s(^2D_{5/2})4p$ & 31.249(2)  & \\
\bottomrule
\end{tabular}
\end{footnotesize}
\end{table}

\section{T\lowercase{ransient} XUV \lowercase{absorption data processing}}\label{sn:data_process}

The transient XUV signal reported in our study was the normalized transmission change given by $\Delta{I}(E,t) = [I_\text{on}(E, t)-I_\text{off}(E)]/I_\text{off}(E)$, where $I_\text{on}(E, t)$ is the transmitted XUV intensity in the presence of pump at delay time $t$, and $I_\text{off}$ was measured when the pump laser was blocked. $\Delta{I}(E,t)$ signals prior to pump pulse arrival were uniformly subtracted for all time delays as here we focused on the femtosecond dynamics instead of the millisecond cumulative heating effects, which do not change the conclusions of this work. To calibrate time zero in the transient time-resolved measurements, we referenced the pump-probe temporal overlap determined from the gas transient (Fig.~\ref{fig:IRF_He}) and then shifted the delay time in the 1$T$-TiSe$_2$ measurements accordingly.

To describe the photoinduced changes observed on a given sample, the temporal traces integrated over certain spectral windows were fit to the following phenomenological model function \cite{Rohwer2011,Zong2021}
\begin{align}
    \Delta{I}(t) = \frac{1}{2}\left[1 + \mathrm{Erf}\left(\frac{2\sqrt{\ln 2}(t-t_0)}{\omega}\right)\right] \cdot\notag \\
    \left(I_\infty + I_0 e^{-(t-t_0)/\tau}\right),\label{eq:fit_erfexp}
\end{align}
where $\omega$ represents the intrinsic system response time, $I_0$ stands for the maximum intensity change, $I_\infty$ is the value of $\Delta I$ at long time delays, $\tau$ denotes the characteristic relaxation time to the quasi-equilibrium, and $t_0$ is associated with the relative arrival time of pump and probe when $\Delta I = (I_0+I_\infty)/2$. To describe the temporal traces with oscillatory phonon features [$\Delta{I_\text{ph}}(t)$], an exponentially-decaying sinusoidal term was added to Eq.~\eqref{eq:fit_erfexp}, leading to 
\begin{align}
    &\Delta{I_\text{ph}}(t) = \frac{1}{2}\Bigg[1 + \mathrm{Erf}\left(\frac{2\sqrt{\ln 2}(t-t_0)}{\omega}\right)\Bigg] \cdot \notag\\
    &\Big[I_\infty + I_0 e^{-(t-t_0)/\tau}+I_\text{ph}e^{-(t-t_\text{ph})/\tau_\text{ph}}\cos(2\pi\nu t+\phi)\Big],\label{eq:fit_erfexp_cos}
\end{align}
where $I_\text{ph}$, $\tau_\text{ph}$, and $\nu$ represent the phonon oscillation amplitude, relaxation time, and frequency, respectively, while $t_\text{ph}$ and $\phi$ are the associated temporal offset and initial oscillation phase.

In the main text, time traces taken from spectral window~iii were fit to Eq.~\eqref{eq:fit_erfexp} because of the absence of coherent phonon oscillations while those from spectral windows~i, ii, and iv were fit to Eq.~\eqref{eq:fit_erfexp_cos}. Due to the additional dynamics arising from the coherent $A_{1g}$ phonon, we hence chose spectral window~iii for analyzing the initial system response at different excitation densities in Fig.~\ref{fig:4}, which shows evidence for excitonic correlations both below and above $T_c$, free from the interference of the $A_{1g}$ phonon. As a control analysis, we show in Fig.~\ref{fig:Phonon_initial_response}a,b the fluence-dependent initial response in spectral windows dominated by the coherent excitation of the $A_{1g}$ phonon. As summarized in Fig.~\ref{fig:Phonon_initial_response}c, the initial response time is independent from the fluence within experimental uncertainty, which is in stark contrast to the anti-correlation between the incident fluence and the initial response time seen in spectral window~iii. The fluence-independent initial response time seen in Fig.~\ref{fig:Phonon_initial_response}c reflects the fact that the frequency of the $A_{1g}$ phonon is independent from the excitation density within the probed range, consistent with the sinusoidal fitting in Eq.~\eqref{eq:fit_erfexp_cos}.

\section{D\lowercase{ensity functional theory calculations}\label{sn:dft}}

To numerically investigate the electronic structures of 1$T$-TiSe$_2$, we employed the open-source package Quantum-Espresso \cite{Giannozzi2017} based on density functional theory (DFT) \cite{Hohenberg1964,Kohn1965}. For practical calculations, we used the Perdew-Burke-Ernzerhof functional \cite{Perdew1996} as an approximation to the exchange-correlation functional. For the description of ions, we utilized the projector augmented-wave method \cite{Blochl1994} with pseudopotentials provided by A.~D.~Corso \cite{DalCorso2014}. For all the calculations performed in this work, we set the cutoff energy for the wavefunction to $55$~Ry and that for the density to $275$~Ry. For the normal phase of 1$T$-TiSe$_2$, we used a unit cell that contains three atoms (one Ti atom and two Se atoms), and the first Brillouin zone is sampled with $24\times 24\times 4$ $k$-points. Here, the third axis is perpendicular to the layer of 1$T$-TiSe$_2$. Similarly, for the CDW phase, we used a unit cell that contains 24 atoms, and the Brillouin zone is sampled by $4\times 4\times 4$ $k$-points. With these conditions, we first computed the ground state by solving the Kohn–Sham equation. Once the self-consistent ground state was obtained, we further analyzed the density of states and the projected density of states by projecting the Kohn–Sham orbitals onto atomic orbitals.

\section{T\lowercase{ime-dependent} DFT \lowercase{calculations} \label{sn:TDDFT}}

We also theoretically investigated the optical properties of 1$T$-TiSe$_2$ using first-principles electron dynamics calculations based on time-dependent density functional theory (TDDFT) \cite{Runge1984}. Here, we briefly describe the theoretical methods used to evaluate the optical properties of solids with TDDFT, while the details of the methods are described elsewhere \cite{Bertsch2000,Sato2021}.

\begin{figure*}[tbh!]
    \includegraphics[scale=0.62]{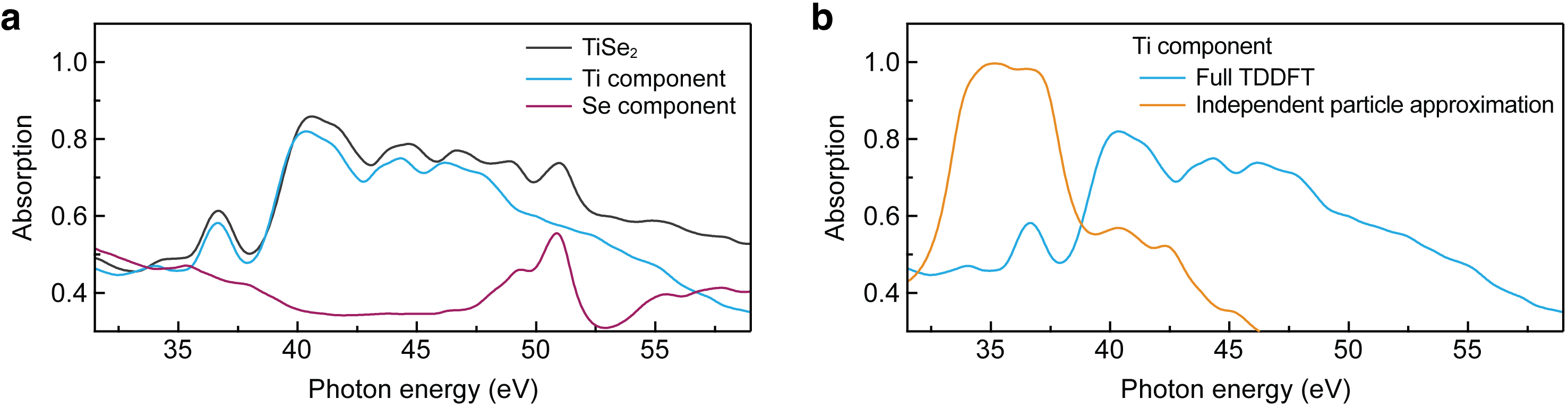}
    \caption{\textbf{XUV absorption spectrum of 1\textit{T}-TiSe$_\text{2}$ computed by TDDFT.} \textbf{a},~Full TDDFT spectrum for 1$T$-TiSe$_2$ in its normal state (black), calculated for a 30-nm-thick sample. Blue and purple curves are contributions from Ti and Se, respectively. Spin-orbit coupling was not considered in the core-level calculation, so only two Se absorption peaks were seen in the theoretical spectrum whereas four peaks were observed experimentally (see Fig.~\ref{fig:1}j,k). \textbf{b},~Comparison of the calculated spectrum from Ti between the full TDDFT result (blue curve, reproduced from \textbf{a}) and under the independent particle approximation (yellow curve). See text for details.}
\label{fig:abs_cal}
\end{figure*}

The dynamics of electronic systems in solids are described by solving the time-dependent Kohn-Sham equation,
\begin{align}
i\hbar \frac{\partial}{\partial t}u_{b\boldsymbol{k}}(\boldsymbol{r},t) = H_{\boldsymbol{k}}(t)u_{b\boldsymbol{k}}(\boldsymbol{r},t),
\label{eq:tdks}
\end{align}
where $u_{b\boldsymbol{k}}(\boldsymbol{r},t)$ is the periodic part of the Bloch wavefunction with the band index $b$ and the Bloch wavevector $\boldsymbol{k}$. Here, the Kohn-Sham Hamiltonian, $H_{\boldsymbol{k}}(t)$, is given by
\begin{align}
H_{\boldsymbol{k}}(t) = \frac{1}{2m}\left[ \boldsymbol{p} + \hbar \boldsymbol{k} + e\boldsymbol{A}(t) \right]^2
+ \hat{v}_{\mathrm{ion}}(t) + \notag\\
v_{\mathrm{H}}(\boldsymbol{r},t) + v_{\mathrm{xc}}(\boldsymbol{r},t),
\end{align}
where $\boldsymbol{A}(t)$ is the spatially uniform vector potential related to applied laser electric fields as $\boldsymbol{A}(t)=\int^t_{-\infty}dt' \boldsymbol{E}(t')$, $\hat{v}_{\mathrm{ion}}$ is the ionic potential, $v_{\mathrm{H}}(\boldsymbol{r},t)$ is the Hartree potential, and $v_{\mathrm{xc}}(\boldsymbol{r},t)$ is the exchange-correlation potential. In this work, we employ the norm-conserving pseudopotential approximation for the ionic potential, $\hat{v}_{\mathrm{ion}}(t)$, by freezing core states of Ti and Se atoms \cite{Oliveira2008}. For the exchange-correlation potential, we use the local density approximation \cite{Perdew1992}.

To evaluate the optical properties of 1$T$-TiSe$_2$, we compute the electron dynamics induced by an impulsive distortion,
\begin{align}
\boldsymbol{E}(t) = k_0 \boldsymbol{e}_p \delta(t),
\end{align}
where $k_0$ is the amplitude of the impulsive distortion and $\boldsymbol{e}_p$ is a unit vector along the $p$-direction. By solving Eq.~(\ref{eq:tdks}) under the impulsive distortion, we obtain the time-evolving Kohn--Sham orbitals, $u_{b\boldsymbol{k}}(\boldsymbol{r},t)$. Furthermore, by using $u_{b\boldsymbol{k}}(\boldsymbol{r},t)$, we compute the electric current induced by the distortion as
\begin{align}
\boldsymbol{J}(t) = -\frac{e}{\Omega} \sum_b \int_{\mathrm{BZ}} d\boldsymbol{k} f_{b\boldsymbol{k}} \int_{\Omega} d \boldsymbol{r} 
u^*_{b\boldsymbol{k}}(\boldsymbol{r},t) \hat{v}_{\boldsymbol{k}}(t) u_{b\boldsymbol{k}}(\boldsymbol{r},t),
\end{align}
where $\Omega$ is the volume of the unit cell, $f_{b\boldsymbol{k}}$ is the occupation factor, and $\hat{v}_{\boldsymbol{k}}(t)$ is the velocity operator defined by $\hat{v}_{\boldsymbol{k}}(t) = \left [\boldsymbol{r}, H_{\boldsymbol{k}}(t) \right ]/i\hbar$.

By applying the Fourier transform to the applied electric field $\boldsymbol{E}(t)$ and the induced current $\boldsymbol{J}(t)$, the optical properties of solids can be evaluated as follows:
\begin{align}
\sigma(\omega) &= \frac{\int^{\infty}_{-\infty} dt \, \boldsymbol{e}_p \cdot \boldsymbol{J}(t) \, e^{i\omega t - \gamma t}}{\int^{\infty}_{-\infty} dt \, \boldsymbol{e}_p \cdot \boldsymbol{E}(t) \, e^{i\omega t - \gamma t}}, \\
\epsilon(\omega) &= 1 + \frac{4\pi i}{\omega} \sigma(\omega),
\end{align}
where $\sigma(\omega)$ is the optical conductivity, $\epsilon(\omega)$ is the dielectric function, and $\gamma$ is the damping parameter, set to $0.5$~eV$/\hbar$.

For practical electron dynamics calculations in this work, we employed the open-source package, \textit{Octopus} \cite{Tancogne-Dejean2020}. To describe 1$T$-TiSe$_2$ in the normal phase, we use a unit cell containing one Ti atom and two Se atoms. In this work, the unit cell is discretized into real-space grid points with a spacing of $0.15$~Bohr, while the first Brillouin zone is discretized into $24\times 24\times 4$ $k$-points. We analyze the contributions of specific semi-core states of Ti and Se to the optical properties by employing two kinds of pseudopotentials for both Ti and Se ions. For Ti, the first type of pseudopotential treats $3s$, $3p$, $3d$, and $4s$ electrons as valence electrons, while the second type of pseudopotential treats $3d$ and $4s$ electrons as valence. Similarly, for Se, the first type of pseudopotential treats $3d$, $4s$, $4p$, and $4d$ electrons as valence, while the second type of pseudopotential treats $4s$, $4p$, and $4d$ electrons as valence. By employing the first type of pseudopotential for both Ti and Se, both Ti~$3p$ and Se~3$d$ semi-core responses are included in the calculations. By contrast, by using the second type of pseudopotentials, we can omit those semi-core contributions. As a result, the contributions from these semi-core states can be analyzed in the optical responses. Figure~\ref{fig:abs_cal}a shows the computed absorption spectra of 1$T$-TiSe$_2$. The black curve represents the computed spectrum with both Ti~3$p$ and Se~3$d$ semi-core states. By contrast, the blue curve depicts the spectrum with Ti~3$p$ semi-core states while freezing Se~3$d$ states, and the purple curve illustrates the spectrum with Se~3$d$ states while freezing Ti~3$p$ states.

By freezing the time-dependence of $v_{\mathrm{H}}(\boldsymbol{r},t)$ and $v_{\mathrm{xc}}(\boldsymbol{r},t)$, the local field effect can also be ignored. We call such an approximation the independent particle approximation. Figure~\ref{fig:abs_cal}b shows the absorption spectra for the Ti component with and without the independent particle approximation. In contrast to the approximate calculation, the full calculation correctly reproduces (i)~the large blue shift of the Ti~$M_{2,3}$ absorption peak relative to the Ti~3$p$ core-level position probed by photoemission (see Fig.~\ref{fig:1}i,j), and (ii)~the presence of a broad many-body continuum peak that extends beyond 50~eV.

\beginSuppRef
\def\bibsection{\section{S\lowercase{upplementary references}}}

\end{document}